\documentclass[
reprint,
superscriptaddress,
amsmath,amssymb,
aps,
floatfix,
]{revtex4-1}
\usepackage{mathrsfs}
\usepackage{graphicx}
\usepackage{dcolumn}
\usepackage{bm}
\usepackage[hidelinks]{hyperref}

\usepackage{natbib}
\usepackage{color}

\newcommand{\vc}[1]{\boldsymbol{#1}}
\newcommand{\Tonset}{T_\text{onset}}
\newcommand{\TMCT}{T_\text{\tiny MCT}}
\newcommand{\TK}{T_\text{\tiny K}}
\newcommand{\TSF}{T_\text{\tiny SF}}
\newcommand{\EIS}{E_\text{\tiny IS}}

\newcommand{\qSF}{q_{12,\text{\tiny SF}}}
\newcommand{\chiea}{\chi_\text{\tiny EA}}
\newcommand{\pspin}{$p$-spin }

\begin{document}

\title{Rethinking mean-field glassy dynamics and its relation with the energy landscape:\\
the awkward case of the spherical mixed \pspin model
}


\author{Giampaolo Folena}
\affiliation{\footnotesize Dipartimento di Fisica, Universit\`{a} ``La Sapienza'', P.le A. Moro 5, 00185, Rome, Italy}
\affiliation{\footnotesize LPTMS, UMR 8626, CNRS, Univ. Paris-Sud, Universit\'e Paris-Saclay, 91405 Orsay, France}
\author{Silvio Franz}
\affiliation{\footnotesize Dipartimento di Fisica, Universit\`{a} ``La Sapienza'', P.le A. Moro 5, 00185, Rome, Italy}
\affiliation{\footnotesize LPTMS, UMR 8626, CNRS, Univ. Paris-Sud, Universit\'e Paris-Saclay, 91405 Orsay, France}
\author{Federico Ricci-Tersenghi}
\affiliation{\footnotesize Dipartimento di Fisica, Universit\`{a} ``La Sapienza'', P.le A. Moro 5, 00185, Rome, Italy}
\affiliation{\footnotesize INFN, Sezione di Roma1, and CNR--Nanotec, Rome unit, P.le A. Moro 5, 00185, Rome, Italy}

\date{\today}

\begin{abstract}
The spherical \pspin model is not only a fundamental model in statistical mechanics of disordered system, but has recently gained popularity since many hard problems in machine learning can be mapped on it. Thus the study of the out of equilibrium dynamics in this model is interesting both for the glass physics and for its implications on algorithms solving NP-hard problems.

We revisit the long-time limit of the out of equilibrium dynamics of mean-field spherical mixed \pspin models. We consider quenches (gradient descent dynamics) starting from initial conditions thermalized at some temperature in the ergodic phase. We perform numerical integration of the dynamical mean-field equations of the model and we find an unexpected dynamical phase transition. Below an onset temperature $\Tonset$, higher than the dynamical transition temperature $\TMCT$, the asymptotic energy goes below the ``threshold energy'' of the dominant marginal minima of the energy function and memory of the initial condition is kept. This behavior, not present in the pure spherical \pspin model, resembles closely the one observed in simulations of glass forming liquids. We then investigate the nature of the asymptotic dynamics, finding an aging solution that relaxes towards deep marginal minima, evolving on a restricted marginal manifold. Careful analysis however rules out simple aging solutions. We compute the constrained complexity in the aim of connecting the asymptotic solution to the energy landscape.
\end{abstract}

\maketitle




\section{Introduction}

Understanding the relation between the dynamical behavior and the underlying energy landscape is a fundamental question in the physics of glassy systems. The same question arises in a broader interdisciplinary perspective in optimization, information theory and computer science. In these fields, 
the comprehension of the performances of classes of iterative algorithms optimizing complicated functions represents both a practical and a fundamental theoretical problem. One notable example, among many, is provided by supervised learning, where one formalizes data learning as the problem of minimizing a loss function that depends parametrically on the data \cite{friedman2001elements}.

In the years, concepts originated from glassy physics have been proven useful in the description of neural networks \cite{amit1992modeling}, protein folding
\cite{frauenfelder1991energy,sfatos1997statistical}, supervised learning 
\cite{seung1992statistical,engel2001statistical}, information and
computational complexity theory
\cite{monasson1999determining,mezard2009information}, and much more.

Fundamental unifying concepts have been provided by theory of Spin Glasses. Its interaction with computer science has produced and refined powerful optimization algorithms, from the ones that aim to mimic physical dynamics, such as Simulated Annealing \cite{kirkpatrick1983optimization}, Parallel Tempering \cite{earl2005parallel} and Quantum Annealing \cite{finnila1994quantum}, to a family of message passing algorithms, such as Survey Propagation \cite{mezard2002analytic} and Approximate Message Passing \cite{donoho2010message}, which are based on the the mean field equations for spin glasses. Finally, Spin Glass theory has provided a collection of models that, disregarding the finite-dimensional embedding of real materials, allow to study glassy phenomena in a simplified setting.

The model that more than any other has shaped our ideas about the relation between landscapes and dynamics has been the spherical $p$-spin model \cite{crisanti_spherical_1992},  a simple system of spherical spins that interact through a disordered and long-range $p$-body Hamiltonian. The model exact solution offers a detailed description for the long time dynamics \cite{crisanti_spherical_1993,cugliandolo_analytical_1993}, the thermodynamics \cite{crisanti_spherical_1992} and the structure of metastable states \cite{crisanti_thouless-anderson-palmer_1995}.  The ensuing physical picture is at the basis of the Random First Order Transition (RFOT) theory \cite{castellani2005spin}.

The spherical $p$-spin model has been used as starting point to understand complex dynamics in several interdisciplinary contexts, for example, it has been proposed as a simplified model to understand the properties of the loss function and the dynamics of learning in deep networks \cite{choromanska2015loss}.
More generally in Machine Learning many fundamental problem involving tensors are NP-hard \cite{HilLim13}, e.g.\ tensor factorization, tensor completion and tensor Principal Component Analysis \cite{MonRic14,BaGhJa18}.
In this class of tensor problems the hardness in solving the problem has been directly connected to the properties of the underlying energy landscape \cite{BeMeMN17,RoBeBC19,biroli2019iron}.
Thus better understanding relaxation in the spherical $p$-spin model has direct consequences on fundamental NP-hard problems in Machine Learning.

A widely used statistical characterization of the energy landscape sampled by the glassy dynamics is the numerical study of the inherent structures (ISs), defined as the local minima of the energy potential reached by a steepest descent procedure \cite{sastry_signatures_1998,sciortino_inherent_1999}. In simulations of glass forming liquids, it is observed that starting from thermalized configurations at temperature $T$, the energy of the corresponding ISs concentrates around a value $\EIS(T)$ that depends on $T$ only below a characteristic onset temperature $\Tonset$. Remarkably, this temperature is higher than the estimated `mode coupling temperature' $\TMCT$ where Mode Coupling Theory would predict structural arrest. Not limited to liquids, a similar effect is also observed in 3D Heisenberg spin glasses \cite{baity2019precursors}. From the algorithmic point of view, this dependence shows the effectiveness of Simulated Annealing: that is, in order to minimize the energy, equilibrating at lower and lower temperatures is a better strategy than a sudden quench from a high energy state.  A similar effect is known in optimization, where a `smart initialization' from carefully chosen initial conditions can greatly improve the quality of the solutions.  A strategy of this kind, e.g.\ a spectral initialization, is often adopted to allow simple algorithms to solve non-convex problems \cite{lu2017phase}. However a general theory based on the energy landscape explaining the effect of a `smart initialization' is lacking. We provide in this work a first comprehensive study of the effect of starting from better than random initial configurations.

According to the common belief, the  temperature dependence of the IS energy is a consequence of activation and should not appear in models with long range interactions: initial conditions thermalized above $\TMCT$ would all have the same value of the IS energy, any memory of the initial condition would be lost after a long enough time, and Simulated Annealing would not give any benefit respect to a sudden quench. This would also mean that in this class of NP-hard problems any smart initialization would be useless. However this belief is based on the solution of the spherical \emph{pure} \pspin model \cite{crisanti_spherical_1992}, which, in a nutshell, gives the following dynamical picture:
(1) From the point of view of equilibrium dynamics, the model provides an exact realization of Mode Coupling Theory scenario. One finds dynamical freezing at a temperature $\TMCT$ where the free-energy does not present singularities \cite{crisanti_spherical_1993}.
(2) Randomly chosen initial conditions evolved according to the Langevin dynamics at temperature $T_f<\TMCT$ fail to equilibrate. They fall in an aging state which fails to attain a time translation invariant long time regime and present non trivial response to perturbations \cite{cugliandolo_analytical_1993}.
(3) Any memory of the initial condition is lost in the aging state, the choice of an initial condition thermalized at temperatures $T_{in} >\TMCT$ would not change the asymptotic of a dynamics in presence of a bath at temperature $T_{f}<\TMCT$. 
There is an universal threshold energy $E_{th}$ that attracts the zero temperature dynamics of any such initial conditions, i.e.\ $E_{IS}(T)=E_{th}$ for all $T>\TMCT$.
(3bis) Initial conditions thermalized at $T<\TMCT$ live in metastable states that can be followed down in temperature \cite{franz_recipes_1995,barrat_dynamics_1996,barrat_temperature_1997}. In that case one reaches an energy $E_0(T)$ which is monotonous in $T$ and smaller than then threshold value $E_{th}$. Unfortunately, such initial conditions cannot be generated in a finite time. 

Such a detailed dynamical picture is deeply rooted in the structure and organization of the stationary points of the energy and the free-energy landscapes \cite{cavagna_stationary_1998}. (A) For $E<E_{th}$ almost all stationary points of the energy are isolated and stable minima. The stability-gap of the minima (i.e.\ the minimum eigenvalue of the Hessian) is uniquely a function of the energy. Saddle points are exponentially suppressed and have only a finite number of unstable directions.  (B) For $E>E_{th}$ almost all stationary points are saddles. (C) Minima with $E=E_{th}$ are marginal. Dynamics is attracted by these threshold states, no matter the effort of thermalization at $T>\TMCT$. It is clear that Simulated Annealing is of no use for such a system. (D) Below-threshold energy minima are separated by extensive barriers. There is a one-to-one correspondence between energy minima and finite temperature metastable states (TAP free-energy minima). Metastable states can be followed up and down in temperature. 

The above general picture, both for the dynamics and for the structure of the landscape, is not restricted to the spherical \pspin model, it is generic to mean-field models with a discontinuous glass transition (one step replica symmetry breaking or 1RSB in spin glass jargon) and has been confirmed by the exact description of glasses of particles in the limit of infinite dimension \cite{charbonneau2014fractal}.  However there are important aspects of the physics of the spherical $p$-spin that are deeply model-specific.
 Its Hamiltonian is homogeneous and does not allow for bifurcations or merging of metastable states while changing the temperature. The energy landscape is simpler than more generic mean-field models.  It is well known that in other mean-field modes like the Ising \pspin model, the Potts glass or even spherical models with mixtures of multi body interactions \cite{barrat_temperature_1997} when metastable states are followed down in temperature bifurcations are encountered. This leads to the glass-to-glass Gardner phase transition that has received a lot of attention in the last years \cite{berthier2019perspective}. 

The study of metastable states below such a transition reveals puzzling aspects. The mean-field solutions describing the evolution of states in temperature may unexpectedly disappear \cite{sun_following_2012} as the temperature is lowered. This anomalous behavior has not received a satisfactory explanation, and its implications remain little studied.  In this paper we re-examine dynamics in models where these phenomena occur and challenge crucial aspects of the \pspin dynamical picture: the emergence of a universal dynamical threshold and the loss of memory of high temperature initial conditions.

The paper is organized as follows. In Sec.~\ref{sec:II} we define of the model. In Sec.~\ref{sec:land} we discuss the properties of the landscape and the number of stationary points of the energy surface of the model. Sec.~\ref{sec:oee} presents the equations describing the off-equilibrium dynamics and the effect of a non-random initial condition. Sec.~\ref{sec:affinities} constitutes the core of the manuscript and discusses the energy of inherent structures, the puzzling properties of the zero temperature dynamics and its relation with the landscape. The results are extended to relaxation in presence of a small thermal noise in Sec.~\ref{app:finite_temp}. We discuss an approximated dynamical solution and the emergence of an onset temperature in Sec.~\ref{sec:approx}. We comment on the results and their consequences in Sec.~\ref{sec:disc}. The paper comprises a large number of appendixes devoted to the discussion of more technical aspects of the work that support and complement the main text. In App.~\ref{app:equil_mixed} we discuss finite temperature dynamics in equilibrium. In App.~\ref{app:num_extrap} we describe the numerical solution of the dynamical equations. App.~\ref{app:resp_short} deals with the properties of the response function at short times in the aging regimes. In App.~\ref{app:agingRSB} we present some formal solutions of the long time asymptotic aging regime and the connections with the effective potential. An approximate solution to the asymptotic dynamics is analysed in App.~\ref{app:empirical}. Finally we show the computation of the constrained complexity of energy minima in App.~\ref{app:counting}.

\section{Model definition}
\label{sec:II}

We concentrate on the mixed \pspin spherical model. This model, while presenting a more generic phenomenology, keeps much of the analytical simplicity of the pure spherical $p$-spin. In thi model states do bifurcate, and the above mentioned anomalous behavior occurs. 
The Hamiltonian of the mixed \pspin model is just a sum of \pspin interacting terms with different $p$ values
 \begin{equation}\label{eq:hamiltonian}
	H_J[\sigma] = -\sum_{p} \sqrt{a_p} \sum_{i_1<...<i_p}^{N}J^{(p)}_{i_1...i_p}\sigma_{i_1}...\sigma_{i_p}
\end{equation}
Choosing the couplings $J^{(p)}_{i_1...i_p}$ as independent Gaussian variables of zero mean and variance $\mathbb{E}(J_p^{i_1...i_p})^2=\frac{p!}{2 N^{p-1}}$, and the coefficients $a_p\ge 0$ with $\sum_p a_p<\infty$, we have that $H_J$ is a random Gaussian function on the $N$-dimensional sphere ($\sum_i \sigma_i^2 = N$) with covariance
 \begin{eqnarray*}
&&\overline{H[\sigma]H[\tau]} = N f(q_{\sigma\tau}) \quad \text{where}\\
&&f(q)=\frac 12\sum_p a_p q^p \quad \text{and}\quad q_{\sigma\tau} = \frac{1}{N}\sum_{i=1}^N \sigma_i\tau_i\;.
\end{eqnarray*}
We choose $f(q)$ as a polynomial in a way that the model has a random first order (RFOT) transition in thermodynamics,  associated to an ideal mode coupling transition in equilibrium dynamics.
The pure \pspin model corresponds to the case of a single monomial $f(q)=\frac12 q^p$; we study a 
more generic class of polynomials $f$ where at least two coefficients are positive (mixed \pspin model).
While we keep the function $f$ generic in formulas, all our numerical results are presented for $f(q)=\frac12 (q^3+q^4)$, and compared 
with the pure case with $p=3$. We checked in other examples, that all the results we get in this particular case are typical of general mixed models. 


To investigate how much of the above picture based on the solution of the pure \pspin model is generic, 
 we perform a high precision integration of the dynamical equations describing the evolution of the system in the thermodynamic limit, starting from a thermalized initial condition at temperature $T$ and subsequently undergoing a gradient descent dynamics. Our main finding is that the above picture is not generic and inhomogeneous models show several unexpected features. The value of the asymptotic energy predicted under the assumption of aging with loss of memory is only found starting from high enough temperatures. We find strong evidence for the existence of an onset temperature $\Tonset>\TMCT$ below which the energy of the IS does depend on the initial temperature $T$ and goes below the threshold value.

\section{The energy landscape} 
\label{sec:land}

In order to put in the right frame our dynamical results, we would like first to discuss the structure and organization of the stationary point of the mixed Hamiltonian. The stationary points of the Hamiltonian $H[\sigma]$ on the sphere $\sum_i \sigma_i^2=N$ verify
\begin{eqnarray}
  \label{eq:2}
 \frac{\partial H[\sigma]}{\partial \sigma_i} +\mu\,\sigma_i = H'_i +\mu\,\sigma_i= 0.
\end{eqnarray}
The parameter $\mu$, that we call radial reaction force, is a Lagrange multiplier that insures the spherical constraint.  
In any stationary point, it takes the value 
\begin{eqnarray}
  \label{eq:3}
   \mu= -\frac{1}{N}\sum_i \sigma_i H'_i.
\end{eqnarray}
The radial reaction is directly related to the nature and the stability of the stationary points, in fact, 
the Hessian matrix (intended to be restricted to fluctuations on the sphere) reads
\begin{eqnarray}
  \label{eq:3b}
  M_{ij}=\frac{\partial H[\sigma]}{\partial \sigma_i\partial \sigma_j}+\mu\,\delta_{ij} = H''_{ij}+\mu\,\delta_{ij}
\end{eqnarray}
It is well known \cite{cavagna_stationary_1998} and
rigorously proven \cite{auffinger_random_2010} that $H''_{ij}$ is a GOE random matrix with variance
$\text{Var}[{H''_{ij}}] = \frac{1}{N}f''( 1)$.  Thus the matrix $M$ has a semicircular
spectral distribution $\rho(\lambda)=\sqrt{(\lambda-\mu)^2-4 f''(1)}/(\pi \sqrt{f''(1)})$ which has a lower band edge at
$\lambda_{min} = \mu - 2 \sqrt{f''(1)}$. 
We notice that the value of the radial reaction determines the nature of the stationary points. \footnote{As well know in random matrix theory, while macroscopic fluctuations of the spectral bulk are suppressed by factors scaling exponentially in $N^2$, the probability of having a finite number of eigenvalues at finite distance from the spectral support is exponentially small in $N$. As in the case of the pure model, given value of $\mu$ such that the spectral support is positive, besides the dominating minima we have saddle points with a finite number of negative directions.}
It is natural to define a marginal value of the radial reaction $\mu_{mg}=2\sqrt{f''(1)}$ that separates minima from saddles with an extensive number of negative directions.
Stationary points of $H$ are overwhelmingly minima if $\mu>\mu_{mg}$ and they are saddles or
maxima if $\mu < \mu_{mg}$. Notice that all marginal minima with vanishing spectral gap lie on the manifold specified by $\mu(\sigma)=\mu_{mg}$. 
In the pure model, in any stationary point the radial reaction and the energy verify $\mu=-p E$. In the mixed model, there is not such a one-to-one relation between $\mu$ and $E$ in stationary points.  This can be seen computing the number of stationary points in (\ref{eq:2}), for which both $E$ and $\mu$ are fixed ${\cal N}(E,\mu)=e^{N\Sigma(E,\mu)}$.

The computation of the complexity $\Sigma(E,\mu)$ was performed in Ref.~\cite{arous_geometry_2018} and we just quote the result here (for completeness, a physicist's derivation and the explicit formula for $D(\mu)$ is provided in Appendix~\ref{app:counting})
\begin{equation}
\label{eq:9}
\Sigma(E,\mu)=D(\mu)-\frac{E^2 f''(1)+(E+\mu)^2 f'(1)
}{2\{ f(1) [f''(1)+f'(1)]- f'(1)^2\}}\;.
\end{equation}
The above expression holds when returns a positive value.

\begin{figure}[t]
	\centering
	\includegraphics[width=\columnwidth]{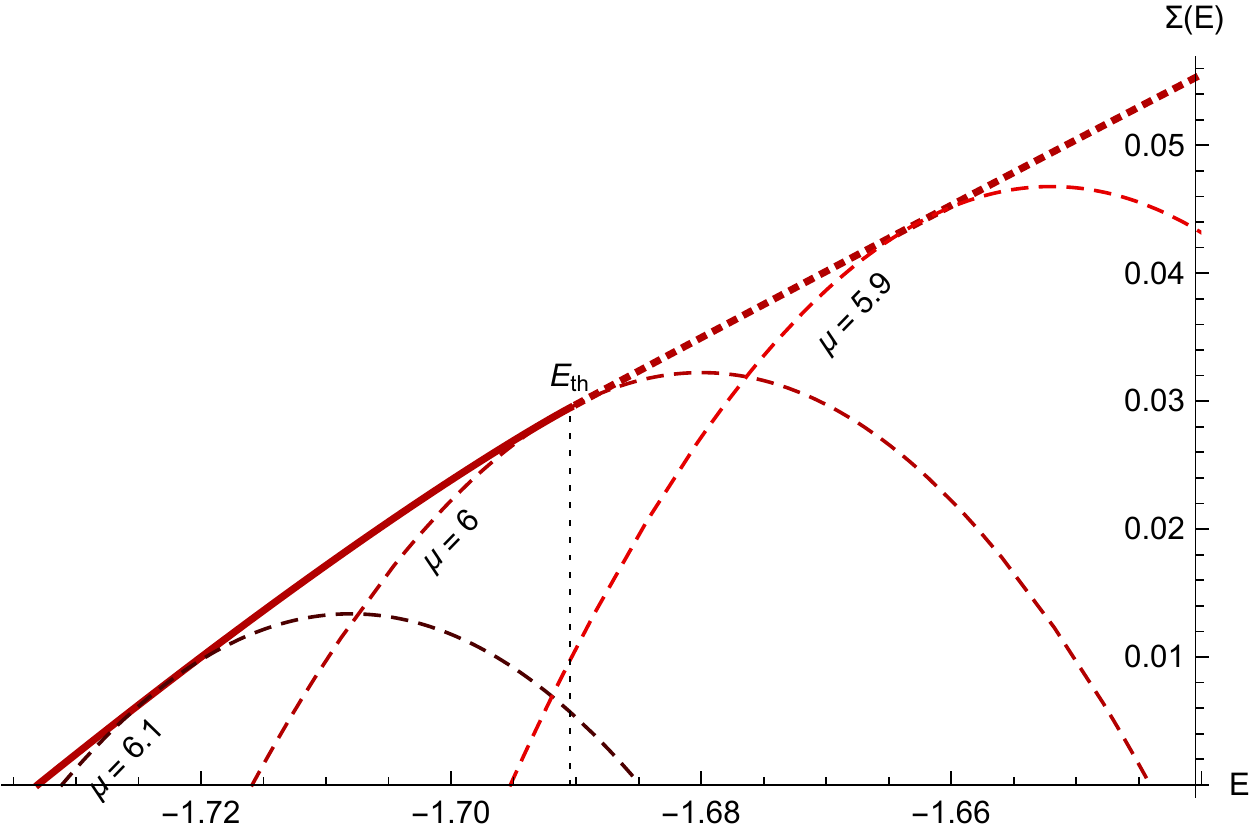}
	\caption{Complexity $\Sigma(E,\mu)$ of the (3+4)-spin model as a function of the energy $E$ for different values of $\mu$. The full line represents the complexity of the dominant stable minima ($\mu>\mu_{mg}=6$). Its continuation above $E_{th}$ represents the complexity of the dominant saddles. The parabolic dashed lines below represent the complexity fixing the radial reaction: $\mu=6$ corresponds to marginal minima, $\mu>6$ to stable minima and $\mu<6$ to saddles. For the same value of $E$ stationary points of different nature coexist.}
	\label{fig:Complexity}
\end{figure} 

Fixing $\mu$, the complexity $\Sigma(E,\mu)$ is a parabola in $E$ with a curvature 
that is finite for mixed models (see Fig.~\ref{fig:Complexity} where we show the complexity for the (3+4)-spin model) and that tends to infinity in the pure limit. For $\mu<\mu_{max}$, with $\mu_{max}$ implicitly defined by equations $0=\Sigma(E_{max},\mu_{max})=\partial_E\Sigma(E_{max},\mu_{max})$, there is a finite interval of energies $[E_-(\mu),E_+(\mu)]$ such that within the interval $\Sigma(E,\mu)>0$. Most importantly, this is true for the threshold value of the radial reaction $\mu_{mg}$. 
Seen from the point of view of $\mu$ the organization of the stationary points resembles the one of the pure model, however from the point of view of the energy there are deep differences. In the pure model, for a given level of energy, the index of any stationary point \footnote{The index counts the number of negative eigenvalues of the Hessian at a stationary point.} differs at most by a finite number from the the index of dominating ones. In the mixed models, for a given energy, stationary points coexist with macroscopically different indices: i.e.\ their spectra are shifted by a finite amount. A unique relation between $E$ and $\mu$ only holds for the exponentially dominating stationary points.

\vspace{8mm}

\section{Dynamical equations in the out-of-equilibrium regime}
\label{sec:oee}

Given the Hamiltonian in Eq.~(\ref{eq:hamiltonian}), 
we consider the following Langevin dynamics
\begin{eqnarray}\label{eq:langevin0}
&&\partial_{t} \sigma_i(t) = -\mu(t) \sigma_i(t)-\frac{\partial H}{\partial  \sigma_i}(t)+\xi_i(t)\\
&&\langle \xi_i(t) \xi_j(t') \rangle = 2T_f\delta_{ij}\delta(t-t')\nonumber\\
&&P[\sigma(0)] = \frac{e^{-\beta H(\sigma(0))}}{Z(\beta)}\nonumber
\end{eqnarray}
where $T_f$ is the temperature of the thermal bath (later set to zero) and $\beta=1/T$ is the inverse temperature at which the initial condition is equilibrated. $\mu(t)$, the time dependent radial reaction force, is the Lagrange multiplier that constrains the dynamics on the sphere, and at time $t$ it takes the value
\begin{eqnarray}
  \label{eq:6}
\begin{small}
  \mu(t)={T_f}-\frac 1 N \sum_i \sigma_i(t)\frac{\partial H}{\partial  \sigma_i}(t).
\end{small}
\end{eqnarray}

Taking the thermodynamic limit, $N\rightarrow\infty$, at finite times, Eq.~(\ref{eq:langevin0}) imply closed integro-differential equations \cite{sompolinsky_relaxational_1982,crisanti_spherical_1993,mezard_spin_1987} for the correlation $C(t,t')\equiv \langle \sigma_i(t) \sigma_i(t') \rangle$ and for the response $R(t,t')\equiv \frac{\partial \langle \sigma_i(t)\rangle}{\partial \xi_i(t')}$. For an initial temperatures $T$ greater than the Kauzmann temperature of thermodynamic phase transition $\TK$ the dynamical equations read \cite{barrat_dynamics_1996}
\begin{widetext}
	\begin{equation}\label{eq:dynamics}
	\begin{aligned}
&	\partial_{t}C(t,t') = -\mu(t)C(t,t') +\int_0^tdsf''(C(t,s))R(t,s)C(s,t')
	+\int_0^{t'}dsf'(C(t,s))R(t',s)+\beta f'(C(t,0))C(t',0)\\
&	\partial_{t}R(t,t') = -\mu(t)R(t,t')+\int_{t'}^t dsf''(C(t,s))R(t,s)R(s,t')\\
&	\mu(t) =  T_f + \int_0^{t}ds f''(C(t,s))R(t,s)C(t,s)+\int_0^{t}ds f'(C(t,s))R(t,s)+\beta f'(C(t,0))C(t,0) \\
&\text{with energy given by}\quad E(t) = -\int_{0}^{t} f'(C(t,s))R(t,s)ds -\beta f(C(t,0))\;.
	\end{aligned}
	\end{equation}
\end{widetext}
The initial condition enters the above equations in a rather simple way. For $\beta=\beta_f$ one gets equilibrium dynamics, whose MCT transition is illustrated in Appendix~\ref{app:equil_mixed}. 
For $\beta=0$ the equations reduce to the usual form valid for uncorrelated initial condition. In this case, for $T_f<\TMCT$ the equations admit a Cugliandolo-Kurchan weak ergodicity breaking aging solution completely analogous to the one of the pure model. It is possible to show that in this solution, for $T_f\to 0$ the radial reaction $\mu(t)$ tends to the marginal value $\mu_{mg}$ for $t\to\infty$, while the energy tends to the threshold value $E_{th}=\frac{-f'(1)^2+f(1)[f'(1)+f''(1)]}{f'(1)\sqrt{f''(1)}}$. In correspondence of this value of the energy, the complexity $\Sigma(E_{th},\mu)$, as a function on $\mu$, is maximum for $\mu=\mu_{mg}$. In other words, $E_{th}$ is not the energy of the most numerous marginal states, but the energy where marginal states exponentially dominate the complexity. 

It is possible to see that for $T<\infty$ the same aging solution solves the dynamical equations in the long time limit, if one assumes that memory of the initial condition is lost and $C(t,0)\to 0$ at a large time. If this would be the correct description for all $T>\TMCT$, then the asymptotic energy would be independent of $T$ in this domain. 

We study the dynamical equations in (\ref{eq:dynamics}) by numerical integration.
The simplest possibility, which we adopt here, is a fixed time step $\Delta t$ first-order Euler discretization algorithm, that gives very reliable results at short times \cite{franz_mean_1994} and allows for reliable extrapolations in the $\Delta t\to 0$ limit (see Appendix \ref{app:num_extrap}). Other algorithms with variable time step have been proposed and used in the literature \cite{kim_dynamics_2001,berthier_spontaneous_2007}. Unfortunately in the case of mixed \pspin model with correlated initial condition ($\beta>0$), these algorithms appear to be unstable at short times and do not allow any improvement over the simplest one. With the Euler integration scheme and a maximum step of $\Delta t=0.1$, we reach times of order $10^3$. These time scales are often not large enough to allow for a simple naive extrapolation to the infinite time limit. Nonetheless, under some conditions, will allow us to make clear claims on the large time dynamics. We concentrate in this paper on the case of zero temperature dynamics, where the energy monotonically decreases and once below threshold it cannot rise up again. We show however in Section \ref{app:finite_temp} that this $T_f=0$ result extends to a range of positive temperatures, $T_f>0$.

\section{Affinities and divergences between mixed and pure models}
\label{sec:affinities}

\begin{figure}[t]
\centering
\includegraphics[width=\columnwidth]{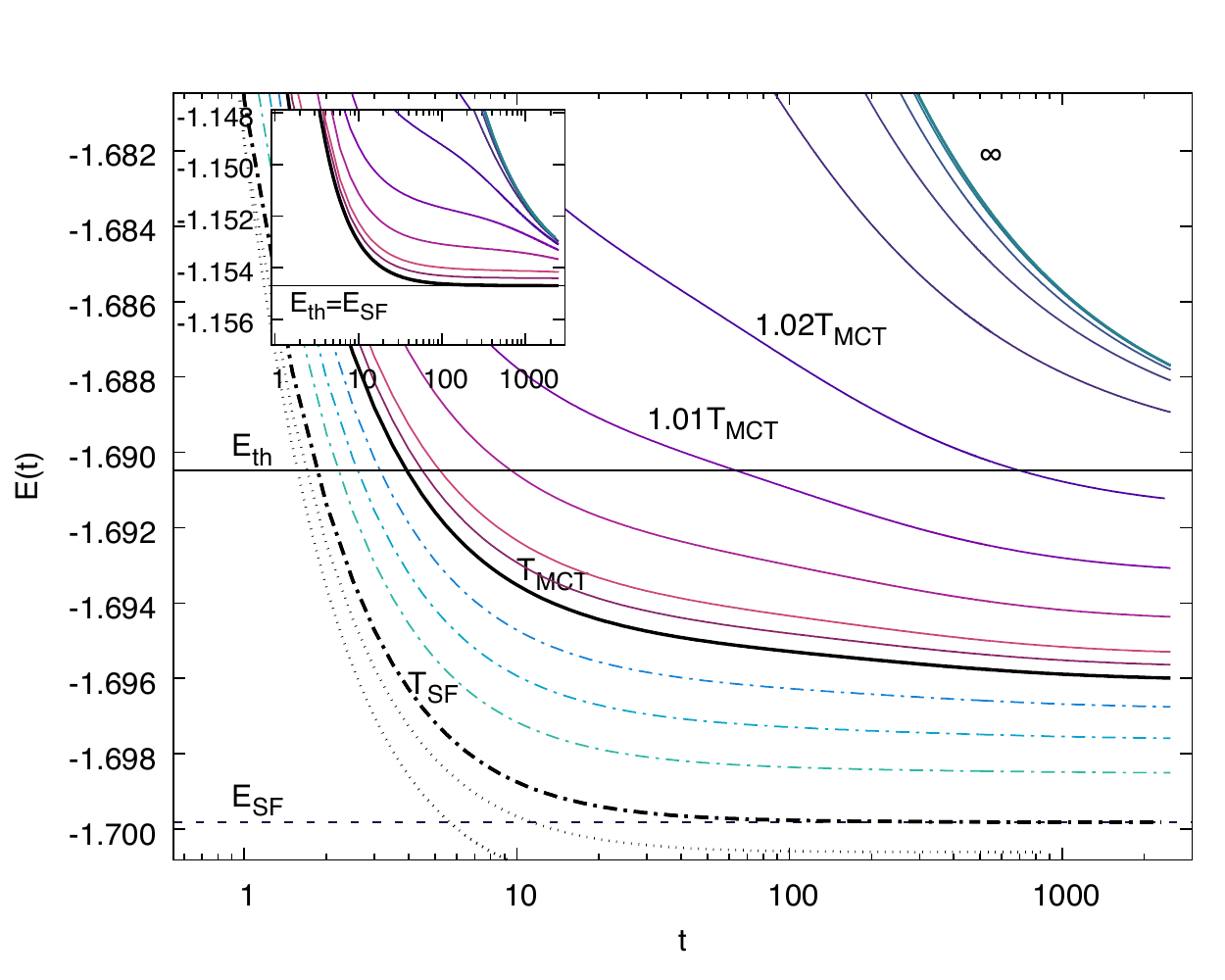}
\caption{Energy relaxation in the (3+4)-spin model, starting from different temperatures $T$ and quenching to zero temperature. Continuous lines are quenches from the ergodic phase ($T\geq \TMCT = 0.805166$), from bottom to top $T/\TMCT=1, 1.001, 1.002, 1.005, 1.01, 1.02, 1.05, 1.1 , 1.2, 1.5, 2, 4, 10, \infty$. Dashed-dotted lines correspond to quenches from temperatures $\TSF=0.9914 \,\TMCT=0.798\leq T<\TMCT$, from top to bottom $T/\TMCT=0.998, 0.996, 0.994, 0.992$ and a last at $\TSF$. Dotted lines correspond to $T<\TSF$ ($T/\TMCT=0.99, 0.988$).  We clearly see that the energy goes below threshold for  temperatures that are small enough, although larger than $\TMCT$. Inset: same dynamics for the 3-spin model ($T\ge \TMCT$), the energy never goes below threshold.} 
\label{fig:Energy}
\end{figure}

We discuss here the core question of this paper, and compare the energy reached by the dynamics from different initial conditions, to the threshold energy that can be computed from the asymptotic solution with $\beta=0$ \cite{cugliandolo_analytical_1993} or by computation of the complexity of minima \cite{crisanti_thouless-anderson-palmer_1995}.  For the 3+4 model the value of the threshold energy is $E_{th}=-71/42\simeq -1.6905$.
In Fig.~\ref{fig:Energy} we show the curves of the energy as a function of time for different $T$, ranging from $T=\infty$ to $T=0.796$, the dynamical transition temperature of the model being $\TMCT=0.805166$ (see Appendix~\ref{app:equil_mixed}). Data are obtained from an integration with step $\Delta t=0.1$, reaching the maximum time of $2500$. This step size gives a relative integration error smaller than $10^{-5}$ (see detailed discussion in the Appendix~\ref{app:num_extrap}).

On the time scale we can reach and for $T\gtrsim 1$ the energy seems to have reached an asymptotic behavior well fitted by a power law $E(t)=E_{th}+a/t^\gamma$ with $a$ slightly dependent on temperature and $\gamma=0.66\pm 0.01$, making us confident that $E_{th}$ is the asymptotic value of the energy in this range of temperatures. On the other hand, if $T$ is small enough but still larger than $\TMCT$, the curves of the energy go below $E_{th}$. For comparison in the upper inset we present the same curves for the pure model with $p=3$. In that case manifestly the energy tends to the threshold value for all $T\ge \TMCT$.  
These results suggest a scenario with a new dynamical transition in the mixed model, with an onset temperature $\Tonset$ separating a memoryless phase for $T>\Tonset$ from a memorious phase at $T<\Tonset$.
It is also interesting to look at $T<\TMCT$. In our data we do not observe  anything special happening at $\TMCT$, it is only below a {\it state following} 
temperature $\TSF\approx 0.798$ that we observe relaxation becoming fast and the energy decaying exponentially to its asymptote. We will see that we 
can predict the asymptotic energy in this region through a quasi-static solution of the dynamical equations (`state following' with memory of the initial 
condition).

We will see in the next section that a small temperature $T_f$ in the dynamical equations does not change this behavior. 


In order to understand the dynamical mechanisms that allow to beat the threshold, we look at the behavior of response and correlations. A first quantity we would like to understand is the correlation with the initial state, $C(t,0)$. According to the usual weak ergodicity breaking scenario this quantity should vanish at large $t$. Both in the mixed and in the pure model we observe that the relaxation of  $C(t,0)$ is slower and slower as $T$ decreases. As for the energy we can identify three temperature regimes for $T>\TK$:
(I) a high temperature regime $T>\Tonset$ where memory of the initial condition is lost and $C(t,0)\to 0$;
(II) an intermediate `hic sunt leones' regime $\TSF<T<\Tonset$  where the relaxation is slow and any extrapolation is difficult; 
(III) a low temperature regime $T<\TSF$ where $C(t,0)$ relaxes exponentially fast to a non zero value $q_{12}$. 

\begin{figure}[t]
	\centering
	\includegraphics[width=\columnwidth]{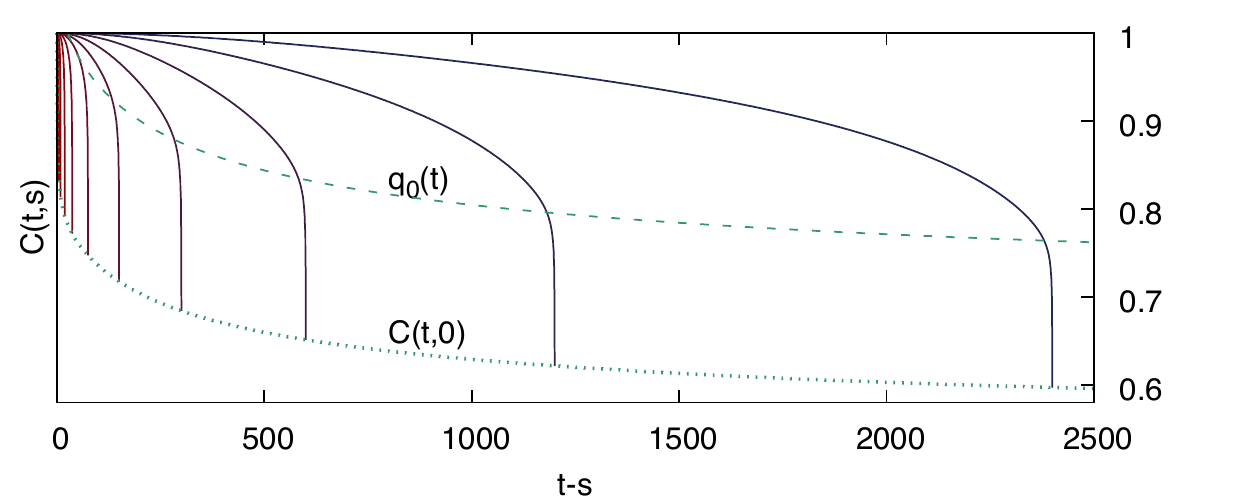}
	\caption{Correlation function $C(t,s)$ for initial temperature $T=0.813=1.01 \TMCT$ as a function of $t-s$ for fixed values of $t=2400\cdot2^{-n}$ with $n=0,\ldots,9$ (from right to left).  The correlation shows clear aging features, becoming slower for larger times. After the slow decay, at a well defined value $q_0(t)$ (dashed line), the correlation display a sudden drop to $C(t,0)$ (dotted line).}
	\label{fig:aging} 
\end{figure}

While no aging is found in regime III, in regimes I and II we find aging, qualitatively similar in the two cases. 
In Fig.~\ref{fig:aging} we show $C(t,s)$ as a function of $t-s$ for several times $t$ and initial temperature $T=1.01 \TMCT$. The curves show a clear aging behavior with the dynamics decorrelating slower and slower as time passes, followed by a fast drop of the correlation for small values of $s$ to the value $C(t,0)$.

\begin{figure}[t]
	\centering
	\includegraphics[width=\columnwidth]{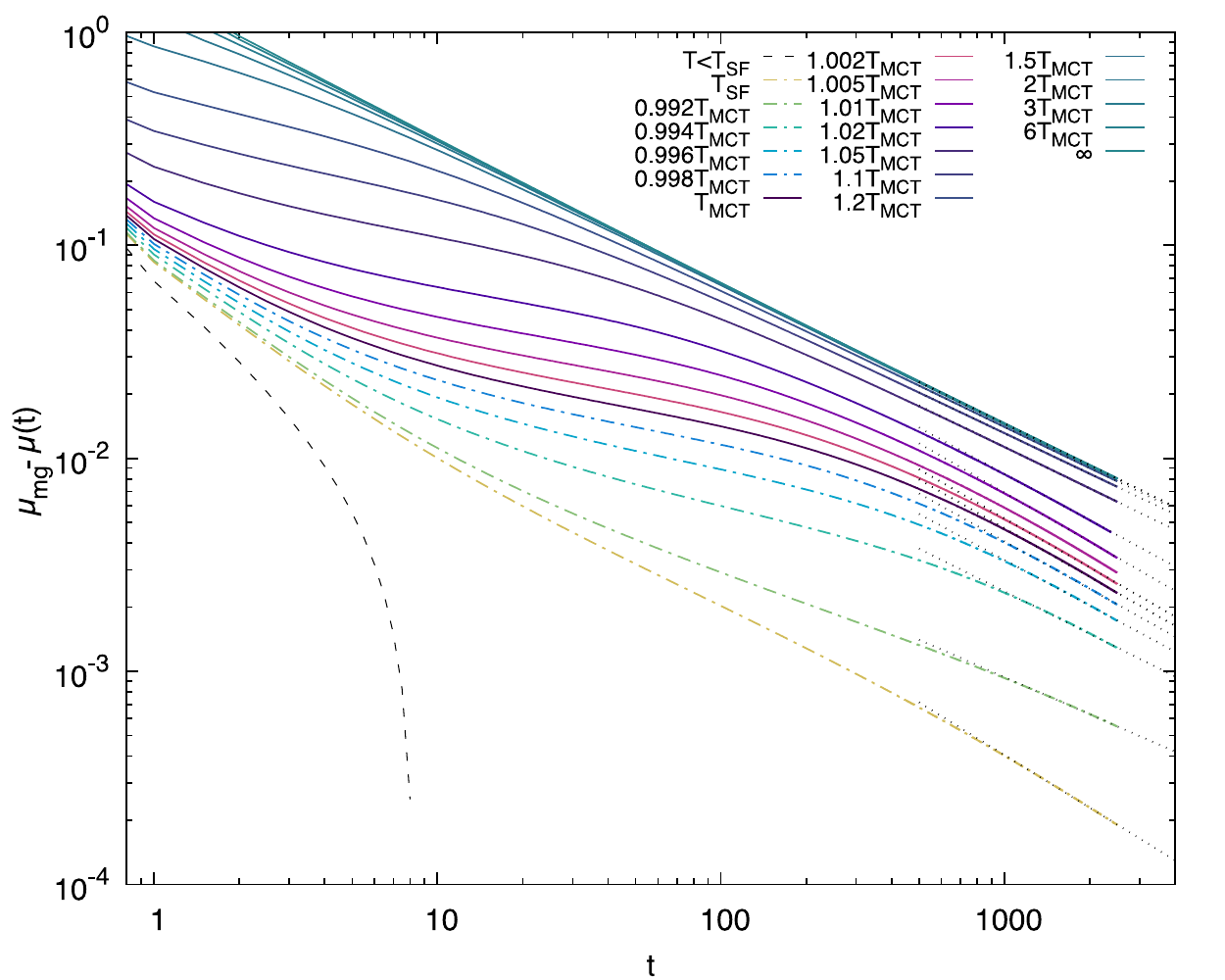}
	\caption{Decay of the radial reaction with time in the (3+4)-spin model on a double logarithmic scale. Its asymptotic value is compatible with $\mu_{mg}=6$ for $T\ge \TSF$ and goes above $\mu_{mg}=6$ for $T< \TSF$. This proves that dynamics tends towards marginal minima for $T>\TSF$.}
	\label{fig:34mufit}
\end{figure}

Despite the energy goes below threshold, we find compelling evidence that the dynamics remain critical, and approaches asymptotically marginally stable minima. 
We reach this conclusion, studying the asymptotic behavior of the radial reaction $\mu(t)$, that for $t\to\infty$ determines the spectral gap of the asymptotically visited states. In Fig.~\ref{fig:34mufit} we can clearly see that both in region I and II, the radial reaction tends to the marginal value $\mu_{mg}\equiv 2\sqrt{f''(1)}$ of marginal minima. In the region III by contrast, the system reaches stable minima and $\mu(\infty)>\mu_{mg}$. 

\begin{figure}[t]
	\centering
	\includegraphics[width=\columnwidth]{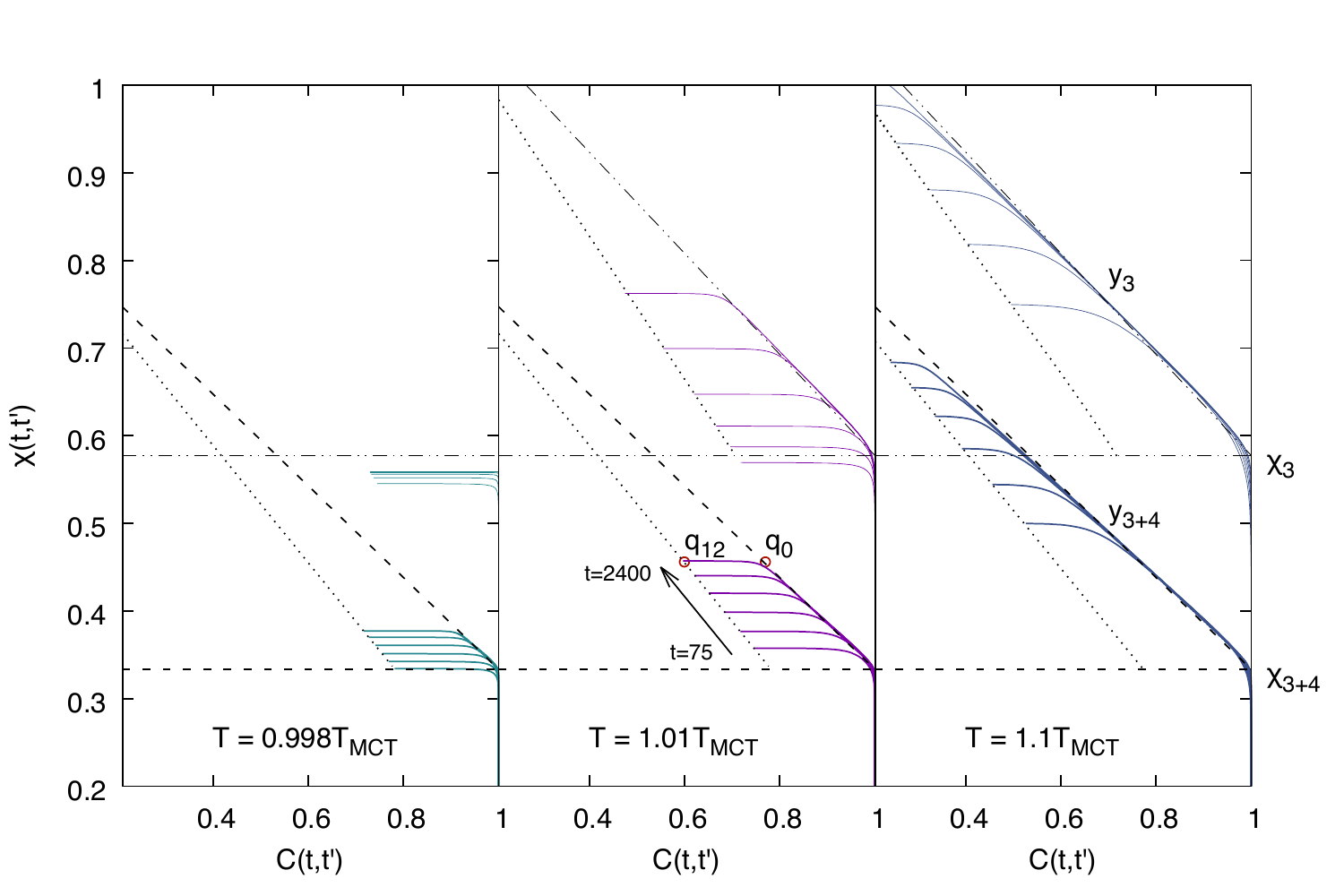}
	\caption{Integrated response $\chi(t,t')$ versus correlation $C(t,t')$ for three different initial temperatures $T/\TMCT=0.998,1.01,1.1$. The lines are plotted fixing (from bottom to top) $t=75,150,300,600,1200,2400$  and changing $t'$. In all the panels upper data are for the 3-spin model and lower data for the (3+4)-spin model. The dashed/dotted lines indicate the susceptibility $\chi_{mg}$ and the fluctuation dissipation ratio $y_0$ predicted by the memoryless solution. It interesting to note that in the (3+4)-spin model the dynamics does not have any appreciable change at $\TMCT$, while in the 3-spin model, starting below $\TMCT$, the system remains trapped in a state.}
	\label{fig:FDR}
\end{figure}

Aging behavior is often qualified studying 
the relation between response and correlation \cite{cugliandolo_energy_1997}.  Fig.~\ref{fig:FDR} shows the parametric plot of the integrated response $\chi(t,s)\equiv\int_s^t R(t,u) \;du$ versus $C(t,s)$ as a function of $s$ for various values of $t$. In zero temperature dynamics, in the large time limit,  the integrated response has a jump in $C=1$ to the intrastate susceptibility $\chiea$. In a given minimum 
\begin{eqnarray}
\chiea=\int d\lambda \frac{\rho(\lambda)}{\lambda}=\frac{\mu+\sqrt{\mu^2-4f''(1)}}{2f''(1)}\;,
\end{eqnarray}
and in marginal minima $\mu=\mu_{mg}$ implies $\chiea=\chi_{mg}\equiv 1/\sqrt{f''(1)}$. 
The curves in Fig.~\ref{fig:FDR} show the formation of a jump in $C=1$ very well compatible with this value. 

The memoryless solution \cite{cugliandolo_analytical_1993} predicts a linear behavior for $C<1$ with slope given by the `fluctuation-dissipation ratio' $y_0\equiv\sqrt{f''(1)}/f'(1)-\chi_{mg}$.  

On the time scales we can access, both the pure and the mixed models show, for all the values of $T>\TSF$, after the jump to $\chi\simeq \chi_{mg}$, an approximately linear part. 
 The linear behaviour continues  
till a rather well defined time dependent value of the correlation $q_0(t)>C(t,0)$ where the response essentially stops to increase.
The value $q_0(t)$ identified in Fig.~\ref{fig:aging} marks the value that separates slow dynamics, where an effective temperature emerges, from fast partial decorrelation from the initial condition, during which the system does not respond (see Appendix~\ref{app:resp_short}). 

Having established that the system relaxes towards mar\-gi\-nal minima, it comes the question of {\it which} mar\-gi\-nal minima are selected. In an attempt to address this question, we study the \emph{constrained complexity} \cite{capone_off-equilibrium_2006} $\Sigma(E,\mu,q_{12};T)$ of energy minima of fixed radial reaction $\mu$, energy $E$ and correlation $q_{12}$ from a reference configuration thermalized at temperature $T$. This quantity for $q_{12}=0$ reduces to the one in Eq.~(\ref{eq:9}) studied in section \ref{sec:land}. The detailed computation of this constrained complexity is presented in appendix \ref{app:counting}.
The computation reveals that for $q_{12}>0$ we have qualitatively the same picture illustrated in Fig.~\ref{fig:Complexity}, but with $\Sigma(E,\mu,q_{12};T)<\Sigma(E,\mu)$, being $\Sigma(E,\mu,q_{12};T)$ monotonically decreasing in $q_{12}$. One can define a generalized threshold energy that depends on $T$ and $q_{12}$, where marginal states dominate the constrained complexity. It turns out that $E_{th}(q_{12},T)$ is a decreasing function of $q_{12}$ and $T$. Unfortunately,  as explained in detail in appendix \ref{app:counting}, while the resulting $E_{th}(q_{12},T)$ has the right qualitative behavior, the corresponding minima have features that are incompatible with the actual attractors of the dynamics. 
The complete static characterization of the attractors of the dynamics remains an important open problem. 

\section{Extending results to finite temperature relaxation dynamics}\label{app:finite_temp}

As we said in section \ref{sec:oee}, for $T_f=0$ the energy monotonically decreases, while if the bath temperature $T_f$ is positive this is not necessarily the case. If we integrate the dynamical equations for a small temperature $T_f$, we observe that the 
finite time energy is continuous, $E(t,T_f)=E(t,T_f=0)+o(T_f)$. We would like to show that,
for small enough $T_f$, this implies that $\lim_{t\to\infty}E(t,T_f)<E_{th}(T_f)$. 

At $T_f>0$ the Lyapunov function of the Langevin dynamics, that decreases monotonically in time, is the free-energy functional \cite{kubo2012statistical}:
\begin{align*}
F(t) &= \int d\sigma P_t[\sigma] H[\sigma] + T_f \int d\sigma P_t[\sigma] \log P_t[\sigma]\\
&\equiv E(t) - T_f S(t)
\end{align*}
where $P_t[\sigma]$ is the probability distribution of configurations at time $t$ induced by the initial condition and the dynamics.

In principle the energy function $E(t)$ could be non-monotonous in time for $T_f>0$, because an entropy increase $\Delta S(t)=S_\infty-S(t)>0$ can induce an energy increase $\Delta E > 0$. However, being the free-energy a decreasing function of time, then we have $\Delta F = \Delta E - T_f \Delta S < 0$, 
and the energy increase is upperbounded 
by 
$\Delta E < T_f \Delta S$. This is a quantity that tends to zero for $T_f\to 0$. Consequently, if the temperature is small enough, 
 the energy $E(t)$, which is below threshold by a $O(1)$ amount, cannot be pushed back to the threshold value.  

\begin{figure}[t]
	\centering
	\includegraphics[width=\columnwidth]{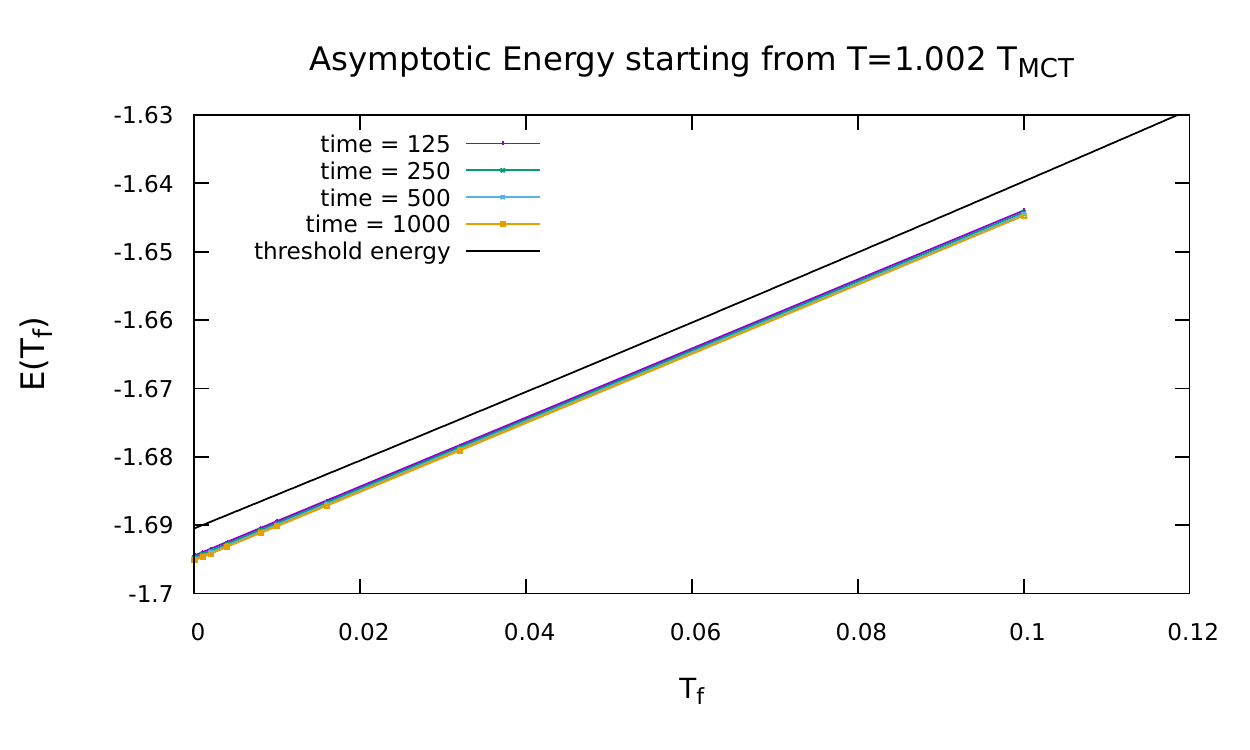}
	\caption{The dependence of the dynamical energy of the (3+4)-spin model as a function of the temperature $T_f$. Data measured at different times are smooth and linear in $T_f$ and clearly below the threshold energy (shown by the upper line). At small enough $T_f$, such an energy difference cannot be compensated by an entropy increase, as discussed in the text.}
	\label{fig:Tpos}
\end{figure}

In Fig.~\ref{fig:Tpos} we show the energy reached at several times $t\in\{125,250,500,1000\}$ during the integration of the dynamical equations as a function of the temperature $T_f$. The topmost full line is the analytic prediction for the threshold energy as a function of $T_f$. We observe that the dynamical energy lies below the threshold energy and shows a continuous and nicely linear behavior in $T_f$. The weak time dependence of the data shown in Fig.~\ref{fig:Tpos} suggests the energy is close to its asymptotic value. Even willing to insist that an eventual entropy increase may induce an energy increase at later times, this increase would be tending to zero for small $T_f$ and thus, by continuity, there would be a finite $T_f$ range where the energy will certainly remain below the threshold value even at infinite time.

\section{An approximate asymptotic solution to the dynamics}
\label{sec:approx}

The dependence of the final energy on $T$ in regime II testifies a form of memory of the initial condition. Differently thermalized configurations lie in basins of attraction of different marginal states. The asymptotic ansatz of Cugliandolo and Kurchan has been  generalized in \cite{barrat_temperature_1997} supposing `aging within a metabasin' and a non vanishing correlation with the initial state $\lim_{t\to\infty}C(t,0)=q_{12}>0$.
Unfortunately, the resulting equations for $\chi,y,q_{12},q_0$ only have an aging solution different from the `amnesic' one ($q_{12}=q_0=0$) in the interval $\TSF\le T\le T_0=0.803 <\TMCT$. Even when the solution exists, the values of the various parameters do not match the observations, in particular one finds $y\approx 0.3$ while in dynamics $y\approx 0.52$. This aging solution terminates at the temperature $\TSF$ where $q_0$ tends to one. Below that temperature, there is a stable `state following' solution with no aging, $q_{12}>0$, $\mu>\mu_{mg}$ and $\chi<\chi_{mg}$ (see Appendix~\ref{app:agingRSB} for a detailed derivation). This solution correctly reproduce the asymptotic values of the energy, correlation, radial reaction of the large time dynamics for temperatures $T \lesssim\TSF$.
Despite all efforts, we have been unable to find a different asymptotic dynamical ansatz describing aging in the `hic sunt leones' regime II and/or to find in the numerics an indication on where the memory of the initial condition could affect the asymptotic solution. The lack of simple aging solutions is certainly related, and equally paradoxical, to the phenomenon of lost of solution in the `state following' procedure of \cite{sun_following_2012}.

\begin{figure}[t]
 	\includegraphics[width=\columnwidth]{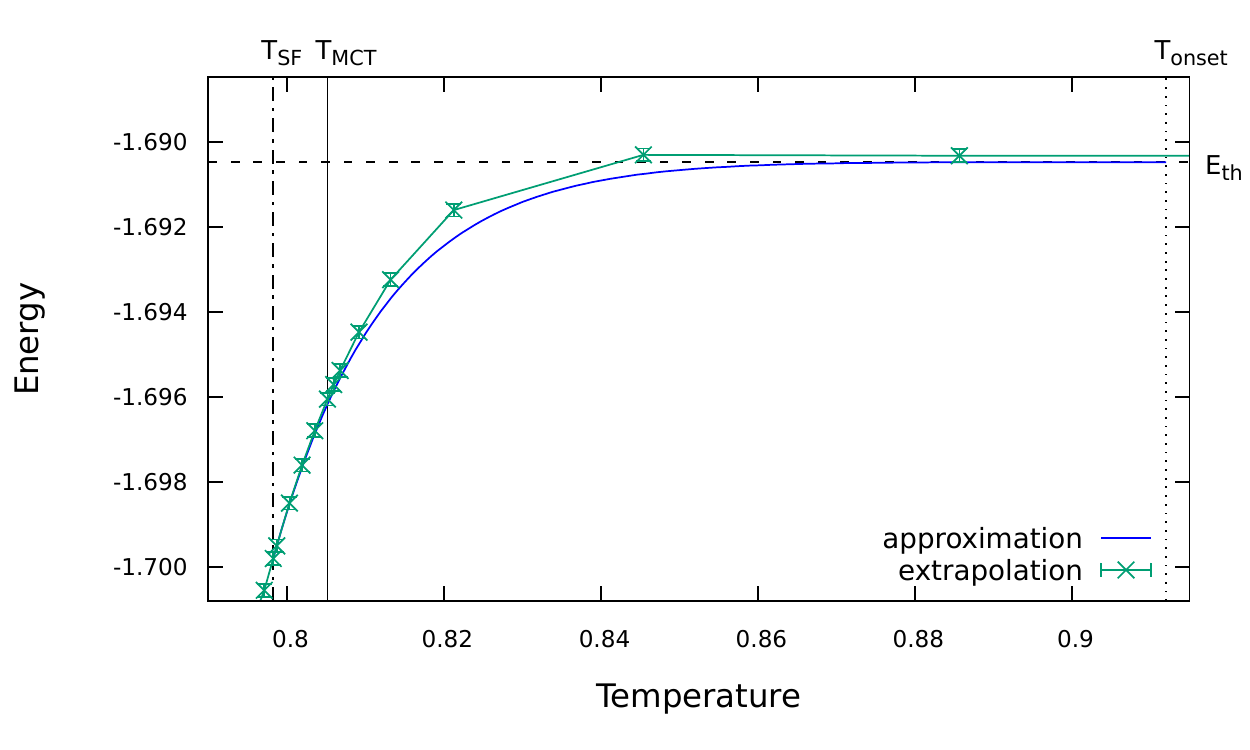}
 	\caption{Numerically extrapolated asymptotic energy (data with errors) and approximated asymptotic solution of dynamical equations (full line) }
 	\label{fig:asympt_energy}
\end{figure}

 Insisting in using a 1RSB dynamical ansatz with $\chiea=\chi_{mg}$, $\mu=\mu_{mg}$ and $y\simeq y_0$, which is supported by the numerical solution, one can easily derive the following relation
\begin{equation}
y_0 f'(q_0) q_0 = \beta f'(q_{12}) q_{12}\;.
\label{eq:q0q12}
\end{equation}
A second relation can be obtained from the observation that $\chi_\text{tot}(t)\equiv\int_0^t ds \;R(t,s) \approx \chi_{mg}+y_0(1-q_0(t))$ depends approximately linearly on $C(t,0)$ even at finite times (see dotted lines in Fig.~\ref{fig:FDR}). Moreover this relation does not present an appreciable temperature dependence in the range $\TSF\le T\lesssim 1$ (data shown in Appendix~\ref{app:empirical}) and implies a linear relation between $q_{12}$ and $q_0$ that can be estimated from two analytically known limits: for $T$ large enough $q_{12}\!=\!q_0\!=\!0$, while at $T\!=\!\TSF$ we have $q_0\!=\!1$ and $\qSF=\sqrt{1-f'(1)/f''(1)}$. Assuming the relation $q_{12}=\qSF\, q_0$, Eq.~(\ref{eq:q0q12}) admits a solution with $q_0>0$ if
$ T<\Tonset\equiv{\qSF^k}/{y_0}$
where $k$ is defined by $f(q)\propto q^k$ for $q\to 0$  (in the 3+4-model $\Tonset=0.91$). This approximated solution gives an asymptotic energy close to the one extrapolated from the numerical integration (see Fig.~\ref{fig:asympt_energy}). Despite this is not an exact solution of the asymptotic equations, it gives us a strong indication that the passage from memoryless to memorious aging could be marked by a phase transition. In addition, it shows that the exact solution cannot be too far from a 1RSB 
weak ergodicity breaking solution inside a basin.


\section{Discussion and perspectives}
\label{sec:disc}
Solving the out of equilibrium relaxation dynamics in the spherical mixed \pspin model, starting from a thermalized configuration at temperature $T$, we have uncovered several interesting and unexpected features.
While the known solution of the relaxation dynamics in the pure \pspin model suggested the existence of a unique dynamical phase transition at $\TMCT$, where ergodicity breaks down, and of a unique threshold energy where the dynamics relaxes below $\TMCT$, our results about the mixed \pspin model reveal a \emph{much richer scenario}.

Our main result is that there is a temperature range above $\TMCT$ where the final energy depends on $T$ and memory of the initial condition is kept. A temperature $\Tonset$ marks the \emph{onset} of this effect.
This resembles realistic glass-formers, but was unexpected in mean field models, because such a phenomenon was ascribed to activated processes, which are absent in the out of equilibrium relaxation of mean-field models. Our results show, contrary to common wisdom, that at $\Tonset$ activated processes do not play any crucial role, and the dependence on the initial configuration can arise even in purely relaxation dynamics. 

From the numerical solution of the dynamical mean-field equations, this dynamical onset of memory may be a cross-over or a new genuine phase transition. We have found an approximate solution of the relaxation dynamics at large times that describes \emph{the onset as a phase transition} and provides, an analytic prediction for the onset temperature.
This provides finally a solvable (although approximate) model whose predictions can be compared with numerical experiments.

Another piece of common wisdom that the present work demystifies is the mantra that `relaxation dynamics in mean-field models relaxes to the threshold energy where the most numerous metastable states are found'.
Calling \emph{threshold energy} the energy where the relaxation dynamics converge if started from a random configuration ($T=\infty$), we have shown that (i) metastable states of infinite lifetime (i.e.\ locally stable states) do exist also above the threshold energy and (ii) the most numerous are above the threshold energy.
The threshold energy 
can be thus understood as the energy value where the measure is dominated by marginal states.
And this can be computed via the \emph{unconstrained complexity}.
Above the threshold energy, an exponential number of marginal states already exist, but the measure over stationary points is dominated by saddles.
The same argument trivializes in the pure \pspin model since in the latter marginal states do exist only at the threshold energy, and so assuming convergence to marginal states determines uniquely the corresponding threshold energy.

We have show evidences that the correct recipe for understanding the large time limit of the relaxation dynamics is to assume that such a dynamical out of equilibrium process always tends to \emph{marginal states}.
For $T<\Tonset$ the persistent memory of the initial configuration suggests to compute a \emph{constrained complexity} of these marginal states that might be correlated with the initial condition. We have performed this task with different levels of accuracy and success.

For an initial temperature below $\TSF$, which is strictly smaller than $\TMCT$ in the mixed \pspin model, the starting configuration lies in a well defined metastable state that can be followed down to zero temperature.
In this situation the dynamics never enters an aging regime and its asymptotic behavior can be computed from a standard \emph{state following} computation.

The most interesting regime is the one reached with an initial temperature $\TSF<T<\TMCT$. The relaxation process falls asymptotically in an aging regime, but remains correlated with the initial configuration.
In this regime, the constrained complexity, while confirming the existence of marginal states below threshold and correlated with the initial condition, fails to identify the ones chosen by the dynamics. 

All our results remain true for dynamics performed at a small temperature $T_f$. 

We believe that, contrary to the pure \pspin model, the mixed \pspin model we studied is generic enough that the properties we uncovered should typically hold in models with a mean-field RFOT, such as Potts glasses \cite{gross1985mean} Ising p-spins \cite{gardner1985spin} and, importantly, glasses of spherical particles in the infinite dimensional limit \cite{charbonneau2014fractal}.


In conclusion, on the basis on the above results, we have to severely revisit our common beliefs about the relaxation dynamics in \pspin models and more generically in models presenting a mean-field RFOT.
The physical picture which has been spread and given credit in our whole scientific community, where a simple connection between the relaxation dynamics and the complexity of metastable states exists, is unfortunately valid only in the case of the pure \pspin model and is false in more general models.

In pure \pspin models strong symmetries lead to the equalities $\TSF=\Tonset=\TMCT$ and these make disappear many of the interesting phenomena that we described in this work and have misled us for two decades with a too simplistic connection between dynamics and energy landscape. Mixed models appear to be well suited to describe the dynamical onset of glassy phenomenology, they have much more complex energy landscapes
which deserve to be better studied in the future. The exact solution to the asymptotic dynamics remains unknown and the connection between asymptotic dynamics and energy landscape is still very open in spherical mixed \pspin models. 
Research is needed in two directions: one should find algorithms capable of reaching large times in solving the dynamical equations and new theoretical ideas on possible structures of the memorious aging are necessary to understand the asymptotic aging regime.  

The richer scenario we have uncovered in this study is likely to have impact also on the many applications in machine learning we discussed in the introduction. Given away the simplest connection between relaxation dynamics and energy landscape, it is clear that more in depth studies of the energy landscape will be needed in order to predict the performance of algorithms (a first example can be found in Ref.~\cite{mannelli2019afraid}), especially of those that do not start from random configuration, but use some smart initialization to improve their performances.

\begin{acknowledgments}
We thank  P.~Charbonneau and G.~Parisi for illuminating discussions. The research has been supported by the Simons Foundation (grants No.~454941, S.~Franz; No.~454949, G.~Parisi), by the European Research Council under the European Unions Horizon 2020 research and innovation programme (grant No. 694925, G. Parisi) and by the project ``Meccanica statistica e complessit\`a'', a research grant funded by PRIN 2015 (Agreement no. 2015K7KK8L). SF is a member of the Institut Universitaire de France. 
\end{acknowledgments}

\appendix

\section{Equilibrium dynamics}
\label{app:equil_mixed}

The onset temperature $\Tonset$ marks the point where landscape influences gradient descent dynamics. In realistic model glasses it has been found that the same temperature marks the onset of non-exponential relaxation, thus making a direct relation between equilibrium dynamics and energy landscape \cite{sastry_signatures_1998}.
It is natural therefore to search for a signature of $\Tonset$ in equilibrium dynamics. 

The aim of this Appendix is to briefly review the  equilibrium dynamics at finite temperature
and show that, while the usual MCT-like transition takes place at $\TMCT$, no anomalies 
around $\Tonset$ are found. The equilibrium dynamical equation for the correlation function reads
\begin{eqnarray}
  \label{eq:7}
  \frac{d C(t)}{dt}=-T C(t)+\beta \int_0^t ds\; f'\big(C(t-s)\big) \frac{d C(s)}{ds}\,.\qquad
\end{eqnarray}
This is readily obtained imposing $T_f=T$ in Eq.~(\ref{eq:dynamics}), together with time translation invariance in the correlation function $C(t,s)=C(t-s)$ and the fluctuation-dissipation relation $R(t)=-\beta  \frac{d C(t)}{dt}$. 
At high temperature, in the ergodic phase, $C(t)$ vanishes at large times. The dynamical MCT-like ergodicity breaking transition is signaled by the appearance of a non zero limit $q_1=\lim_{t\to\infty} C(t)$, with $q_1$ discontinuous at the transition. 
Evaluating  (\ref{eq:7}) at large time, one finds that $q_1$ is solution of 
\begin{eqnarray}
  \label{eq:10}
  \beta^2 f'(q_1)(1-q_1)=q_1. 
\end{eqnarray}
This equation coincides with the equation defining the non-trivial minimum of the equal-temperature effective potential discussed in detail in Appendix \ref{app:agingRSB}.

\begin{figure}[t]
	\centering
	\includegraphics[width=0.95\columnwidth]{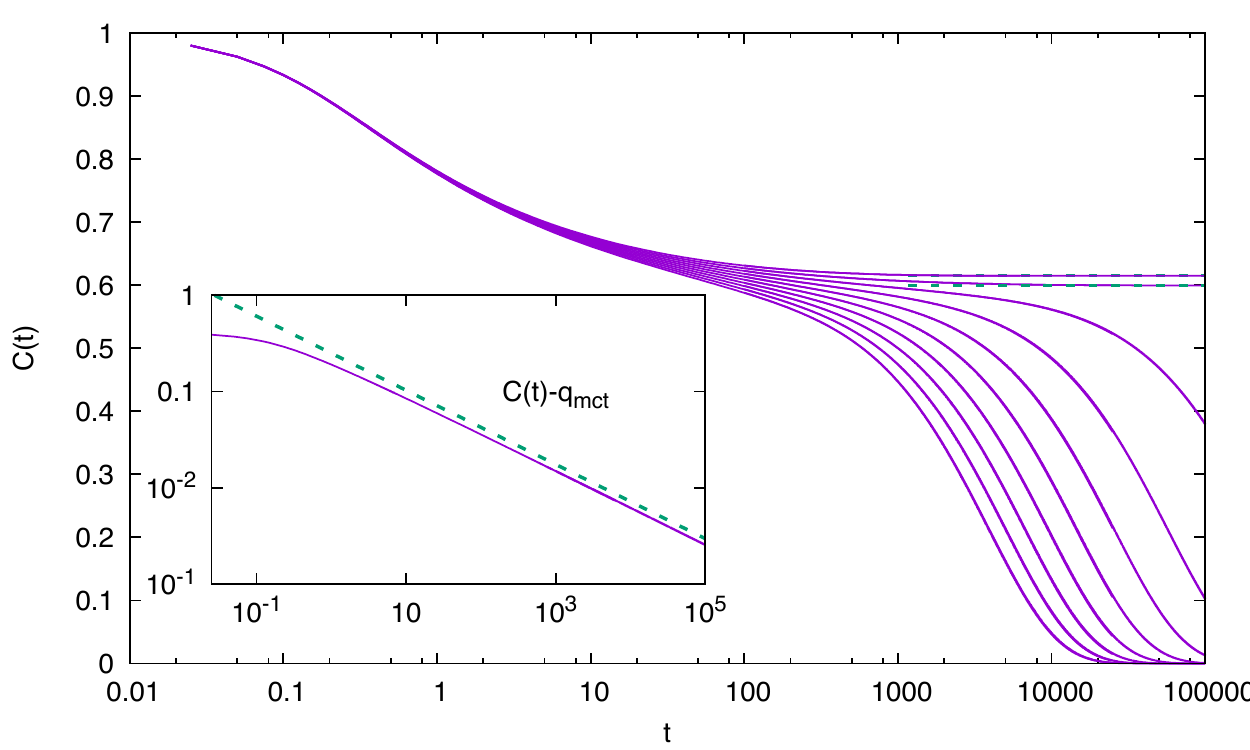}
	\caption{Main panel: Equilibrium correlation function as a function of time on a log-linear scale at different temperatures from $T=0.804$ to $T=0.813$ in steps $\Delta T=0.001$ (from top to bottom).
    All but the first two curves are for $T>\TMCT=0.805166$. We clearly see the formation of a plateau as $\TMCT$ is approached from above.
    The two curves below $\TMCT$ correspond respectively to $q_1=0.599574$ and $q_1=0.614868$, which we draw with dashed lines. Inset: The correlation $C(t)-q_1$ at the critical temperature $\TMCT$ on a log-log scale. For comparison the dashed line represents the expected behavior, a power law with exponent $a=-0.38504$.}
	\label{fig:equilibrium} 
\end{figure}

The transition temperature, the highest temperature for which Eq.~(\ref{eq:10}) has a non zero solution, satisfies
$\beta^2 f''(q_1)(1-q_1)^2=1$.
Standard mode coupling analysis of the dynamics \cite{gotze2008complex} describes the approach to the transition in terms of the formation of a plateau in the correlation function as a function of log-time. This analysis in particular shows that the approach to $q_1$  at $\TMCT$ follows a power law with an exponent $a$ which is determined by $\Gamma(1 - a)^2/\Gamma(1 - 2 a)=\frac{T}{2} f'''[q]/f''[q]^{3/2}$.  For the mixed model $f(q)=\frac 12 (q^3+q^4)$, the transition temperature is $\TMCT=21/\big(2\sqrt{37 \sqrt{37}-55}\big)\approx 0.805166$, corresponding to $q_1=\big(1+\sqrt{37}\big)/12\approx 0.59023$ and $a=-0.38504$. In Fig.~\ref{fig:equilibrium} we display data from the numerical integration of (\ref{eq:7}) with a simple first order Euler algorithm similar to the one we use in the off-equilibrium case. The curves show that theoretical expectations are met: a transition occurs at $\TMCT$ according to the expected pattern and below $\TMCT$ the system is in a non-ergodic state.

\begin{figure}[t]
	\centering
	\includegraphics[width=0.9\columnwidth]{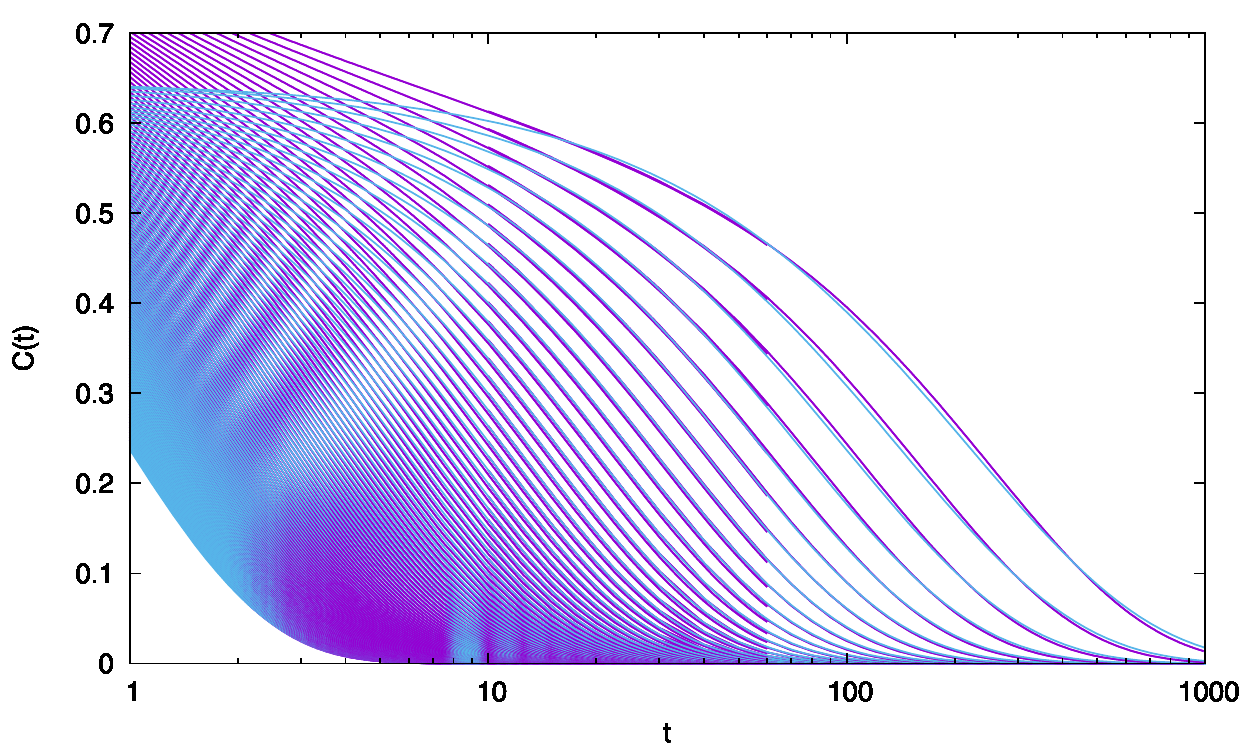}
	\caption{Stretched exponential fit to the equilibrium correlation function $C(t)$ for temperatures in the range $T\in[0.84,2]$ in steps $\Delta T=0.01$. The fit has acceptable quality if the temperature is not too close to $\TMCT$.}
	\label{fig:fit34} 
\end{figure}

\begin{figure}[t]
	\centering
	\includegraphics[width=0.9\columnwidth]{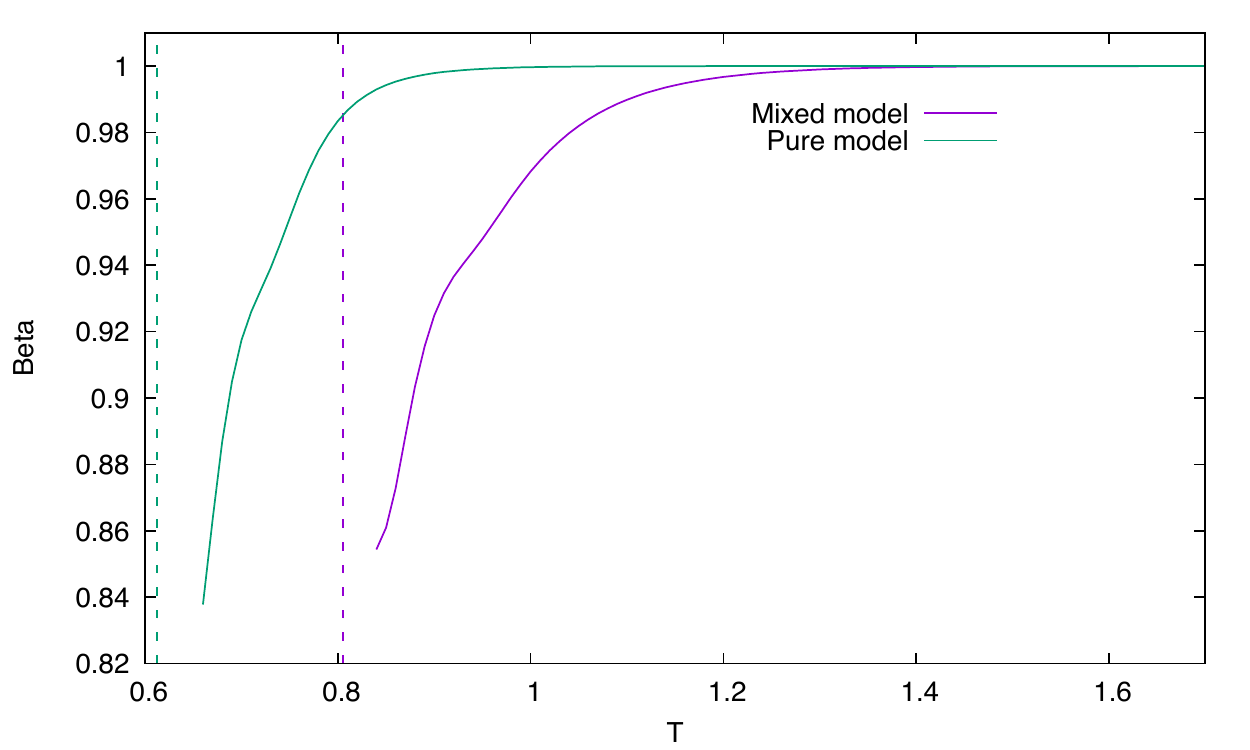}
	\caption{The stretching exponent $\hat{\beta}$ as a function of temperature in the mixed (3+4)-spin model and in the pure 3-spin model. The vertical lines correspond to the transition temperatures $\TMCT$ of the two models. We see no qualitative differences between the two cases. Moreover 
	in the case of the mixed model deviations from pure exponential behavior are seen much above the onset temperature $\Tonset \simeq 0.91$.}
	\label{fig:beta} 
\end{figure}

We then investigate the temperature range around $\Tonset$ to see if this temperature corresponds to a qualitative change in the equilibrium dynamics. Following \cite{sastry_signatures_1998}, we fit the correlation function at intermediate times with a stretched exponential $f(t)=a \exp\big(-(t/\tau)^{\hat\beta}\big)$ and study the behavior of the stretching exponent $\hat\beta$ as a function of the temperature \footnote{We use $\hat\beta$ instead of the standard symbol $\beta$ to avoid confusion with the inverse temperature.}. In \cite{sastry_signatures_1998}, for Lennard-Jones liquids, $\hat\beta$ was shown to become sensibly different from 1 around $\Tonset$.  While the long time behavior of the correlation function is a pure exponential, we can fit the decay at intermediate times. 
In Fig.~\ref{fig:fit34} we see that the stretched exponential form provides a decent fit if we do not approach too much close $\TMCT$.

\begin{figure*}[ht]
	\centering
	\includegraphics[width=\textwidth]{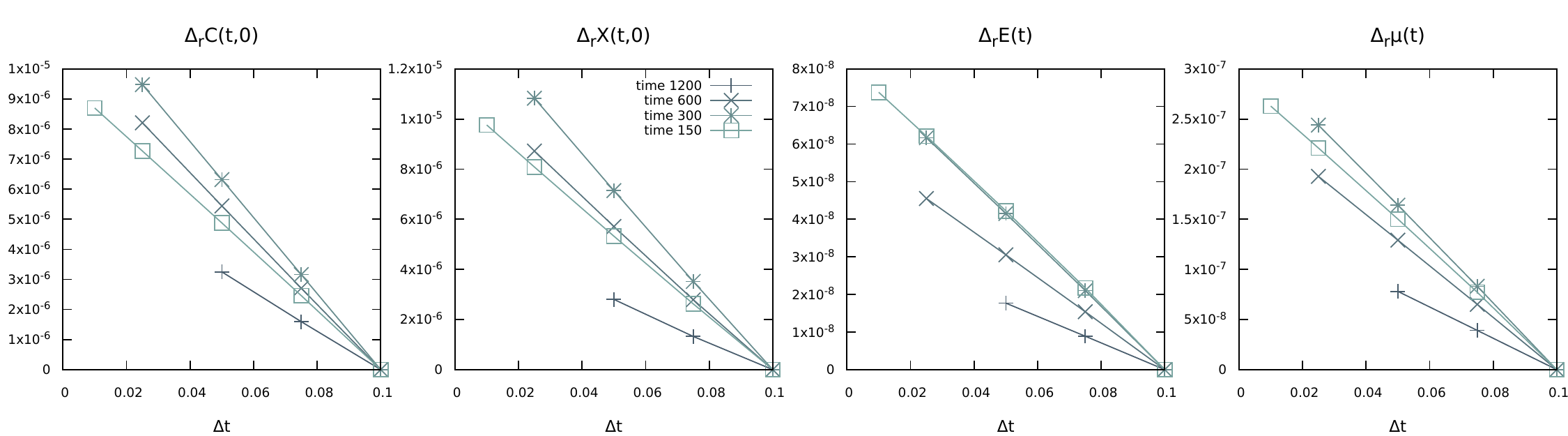}
	\caption{The relative errors, as defined in Eq.~(\ref{eq:relErr}), for $C(t,0)$, $\chi(t,0)$, $E(t)$ and $\mu(t)$ obtained in the integration of the dynamical equations for the (3+4)-spin with $T=\TMCT$. The dependence is linear in the integration step $\Delta t$ and the resulting integration error in using $\Delta t=0.1$ is very small compared to the precision needed in the asymptotic evaluation of the observables.}
	\label{fig:extrap} 
\end{figure*}

We plot the stretching exponent $\hat{\beta}$ for the mixed model as a function of the temperature in Fig.~\ref{fig:beta}, together with the same exponent in the pure model. In both cases we observe that while $\hat\beta\approx 1$ at high enough temperature and it starts to decrease from that value as temperature is lowered. 
We could not find any qualitative difference here in the behavior of the pure and of the mixed model, and we observe that in the mixed model the stretching starts for $T\simeq 1.2$, well above the estimated value of $\Tonset \simeq 0.91$.
Concluding, we do not see any signature of $\Tonset$ in the equilibrium dynamics of the mixed model.

\section{Numerical extrapolations}
\label{app:num_extrap}

The dynamical equations for the correlation and the response functions, written in Eq.~(4), were integrated via a simple Euler algorithm with a fixed integration step $\Delta t$. In this way we get extremely precise results at finite time, but the times we can reach are limited. We tried more sophisticated integration schemes \cite{kim_dynamics_2001}, where the integration steps is increased during the evolution but unfortunately,  as soon as a mixture is used we have met numerical instabilities at very short times. In practice only the pure \pspin model seems to allow those integration schemes to work.

The results presented in the main text have been obtained integrating with an optimal integration step $\Delta t=0.1$, which allows us to reach the largest times without facing any numerical instability (consider that at short times the differential equations we are solving have a natural times scale of order 1, and so $\Delta t$ cannot be much larger than the value we used).

The integration error in the Euler algorithm is linear in the integration step $\Delta t$, and we are interested in understanding how much physical quantities computed with $\Delta t=0.1$ do differ from the corresponding $\Delta t\to 0$ limit.
To this purpose we study the relative errors in one-time quantities defined as follows
\begin{equation}
\Delta_r x(t) = \frac{x(t;\Delta t) - x(t; \Delta t=0.1)}{x(t;\Delta t)}\;.
\label{eq:relErr}
\end{equation}
By definition this relative error is zero for $\Delta t=0.1$, should be linear in $\Delta t$ for $\Delta t$ small enough and in the limit $\Delta t\to 0$ provides the relative error committed in using an integration step $\Delta t=0.1$ instead of the exact integration ($\Delta t\to 0$).

\begin{figure}[t]
	\centering
	\includegraphics[width=0.9\columnwidth]{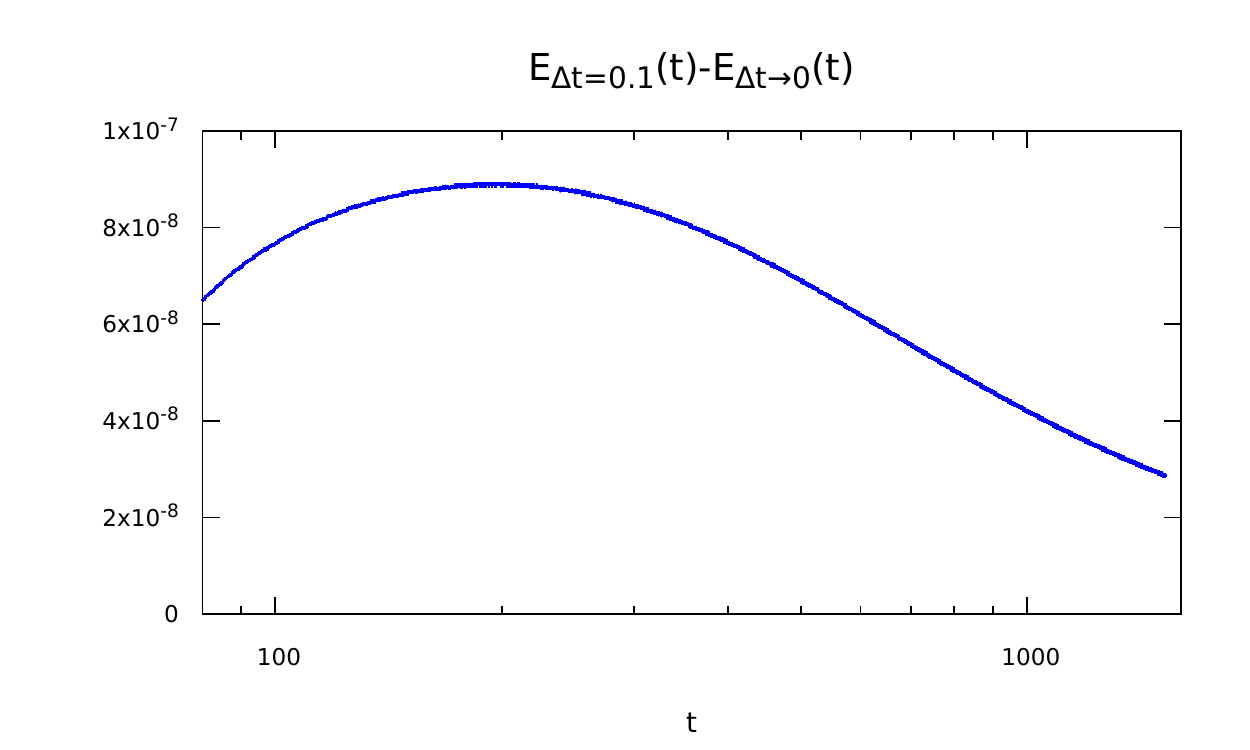}
	\caption{Relative errors in the integration of the dynamical equations for the (3+4)-spin show a maximum at a finite time. Here we plot the relative error in $E(t)$ during the integration with $T=\TMCT$. We notice that the maximum relative error is much smaller, by several orders of magnitude, than the physical effect we seek for (energy going under the threshold).}
	\label{fig:time_plot} 
\end{figure}

In Fig.~\ref{fig:extrap} we show the relative errors in some one-time quantities, $C(t,0)$, $\chi(t,0)$, $E(t)$ and $\mu(t)$, obtained in the integration of the dynamical equations for the (3+4)-spin with $T=\TMCT$. We clearly see that at any time $t$ the relative error scales roughly linear with $\Delta t$. For any time $t$ the relative errors are extremely small: of order $10^{-5}$ for correlation and response, and of order $10^{-7}$ for energy and radial reaction. Moreover the coefficient of the linear relation does not seem to grow indefinitely with time, but actually show a maximum around $t=O(100)$.

In order to better appreciate the non monotonic behavior of the relative errors, we plot in Fig.~\ref{fig:time_plot} the integration error in the energy obtained starting from $T=\TMCT$ as a function of time.
The fact that relative errors decrease at very large times is probably related to another important observation: trajectories integrated with different $\Delta t$ may differ sensibly at short times, but at large times they seem to converge to the same asymptotic solution, which thus are very attractive and stable (this may be the reason why integration errors at very large times are smaller than those at intermediate times). The results of the integration shown in the present work are very reliable and stable.  

Despite the high precision at finite times, the extrapolation to large $t$ of physical quantities, notably $\mu(t)$, $E(t)$ or $C(t,0)$ is delicate. While $C(t,0)$ decays very slow and we do not attempt any extrapolation in the $t\to\infty$ limit,  both $\mu(t)$ and $E(t)$ converge fast enough and their large time limits can be estimated.

We already noticed in Fig.~\ref{fig:34mufit} that for $T\ge\TSF$ the radial reaction $\mu(t)$ is perfectly compatible with an extrapolation to the marginal value $\mu_{mg}=6$. At very high temperatures the decay is a simple power law, while at lower temperatures we observe pre-asymptotic effects, and an eventual crossover to the asymptotic decay. The latter seems to be characterized by an exponent which is roughly temperature independent.

\begin{figure}[t]
	\centering
	\includegraphics[width=0.9\columnwidth]{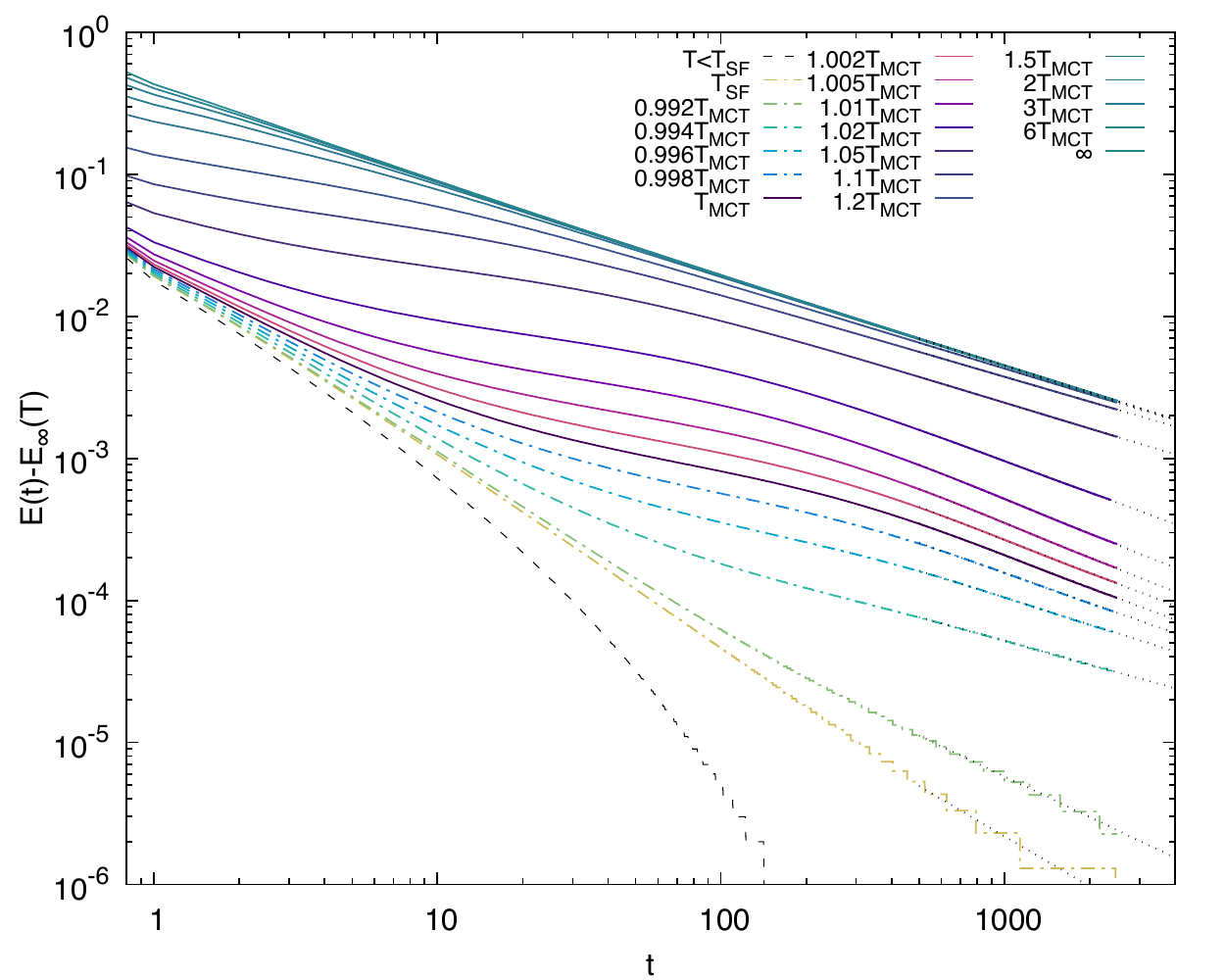}
	\caption{Energy decay with time in the (3+4)-spin model on a double logarithmic scale. The asymptotic value $E_\infty(T)$ has been estimated via power law fits, that provide a good description of the asymptotic decay for $T\ge \TSF$.}
	\label{fig:34Efit}
\end{figure}

Given the critical character of the dynamics in the aging regimes I and II, we have estimated the asymptotic energy $E_\infty(T)$ via power law fits to $E(t)$ data and we show in Fig.~\ref{fig:34Efit} the results. While at high temperatures the data follow a nice power law in time for about a couple of decades, at lower temperatures the presence of the pre-asymptotic regime and the crossover to the asymptotic one, makes very hard to assess the reliablity of the value of $E_\infty(T)$ that we extrapolate in this way.

\begin{figure}[t]
	\centering
    \includegraphics[width=\columnwidth]{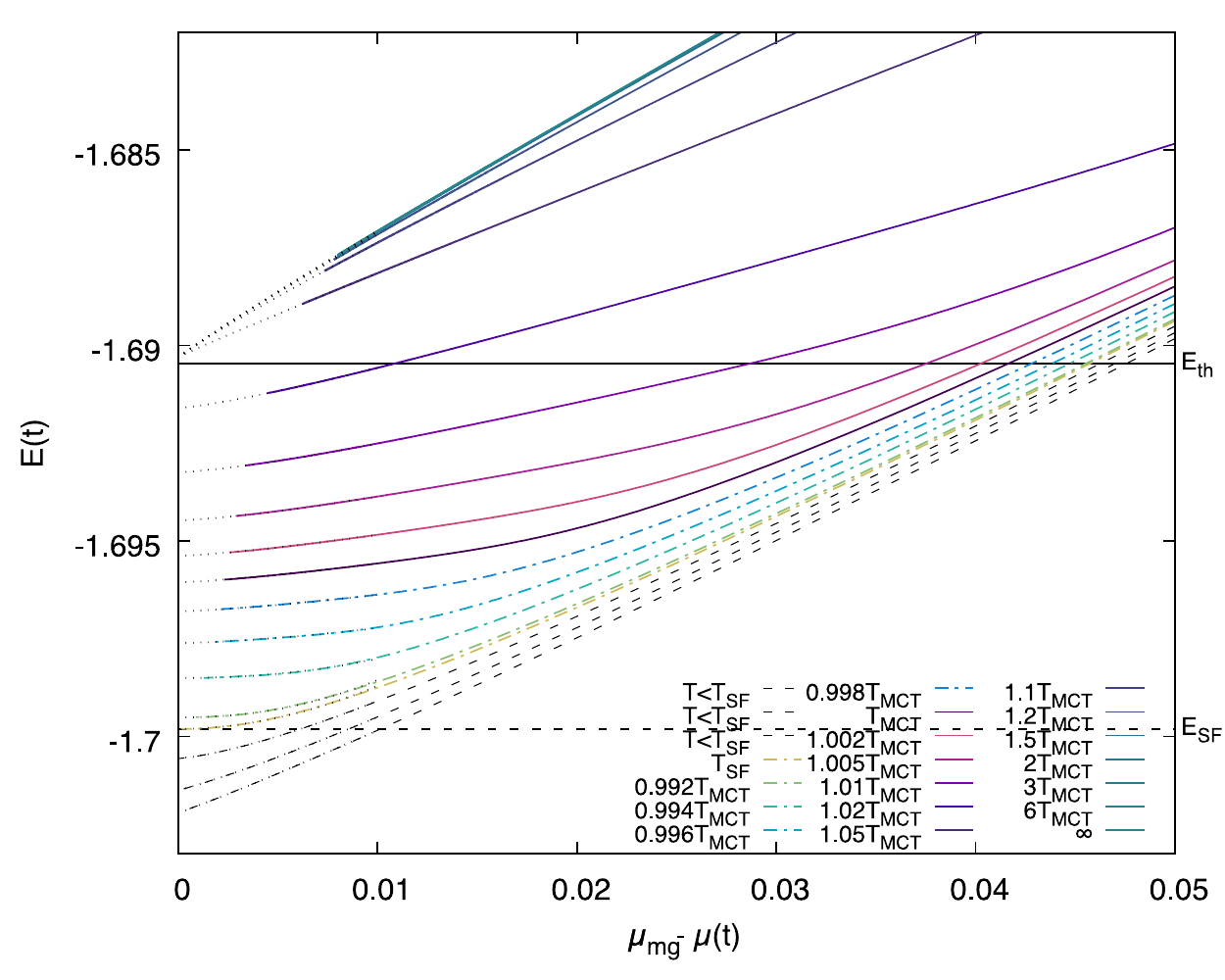}
	\caption{The parametric plot of $E(t)$ versus $\mu_{mg}-\mu(t)$ varying $t$ allows for more reliable extrapolations. The dotted lines are the best power law fits in the form $E=E_\infty(T)+A(T)(\mu_{mg}-\mu)^{\alpha(T)}$ used to estimate the asymptotic energies.}
	\label{fig:34Emu}
\end{figure}

However, comparing the data shown in Fig.~\ref{fig:34mufit} for $\mu(t)$ and in Fig.~\ref{fig:34Efit} for $E(t)$ we notice a very similar behavior (including the crossover) and thus it could be useful to plot parametrically $E(t)$ versus $\mu(t)$ varying $t$, and see whether a better estimation of $E_\infty(T)$ could be obtained. 
This is done in Fig.~\ref{fig:34Emu}. The behavior of the energy as a function of the radial reaction is very smooth, practically linear at high temperatures and close to quadratic approaching $\TSF$. We can then fit the relation $E(\mu)$ and 
obtain a very good estimate of $E_\infty(T)$ just assuming that $\mu(t)$ converges to the marginal value $\mu_{mg}$ in the large time limit. We have interpolated the data via the function
\begin{equation}
E(\mu) = E_\infty + A\, (\mu_{mg} - \mu)^\alpha\;,
\end{equation}
reporting with dotted curves in Fig.~\ref{fig:34Emu} the best fits.

\begin{figure}[t]
	\centering
	\includegraphics[width=0.9\columnwidth]{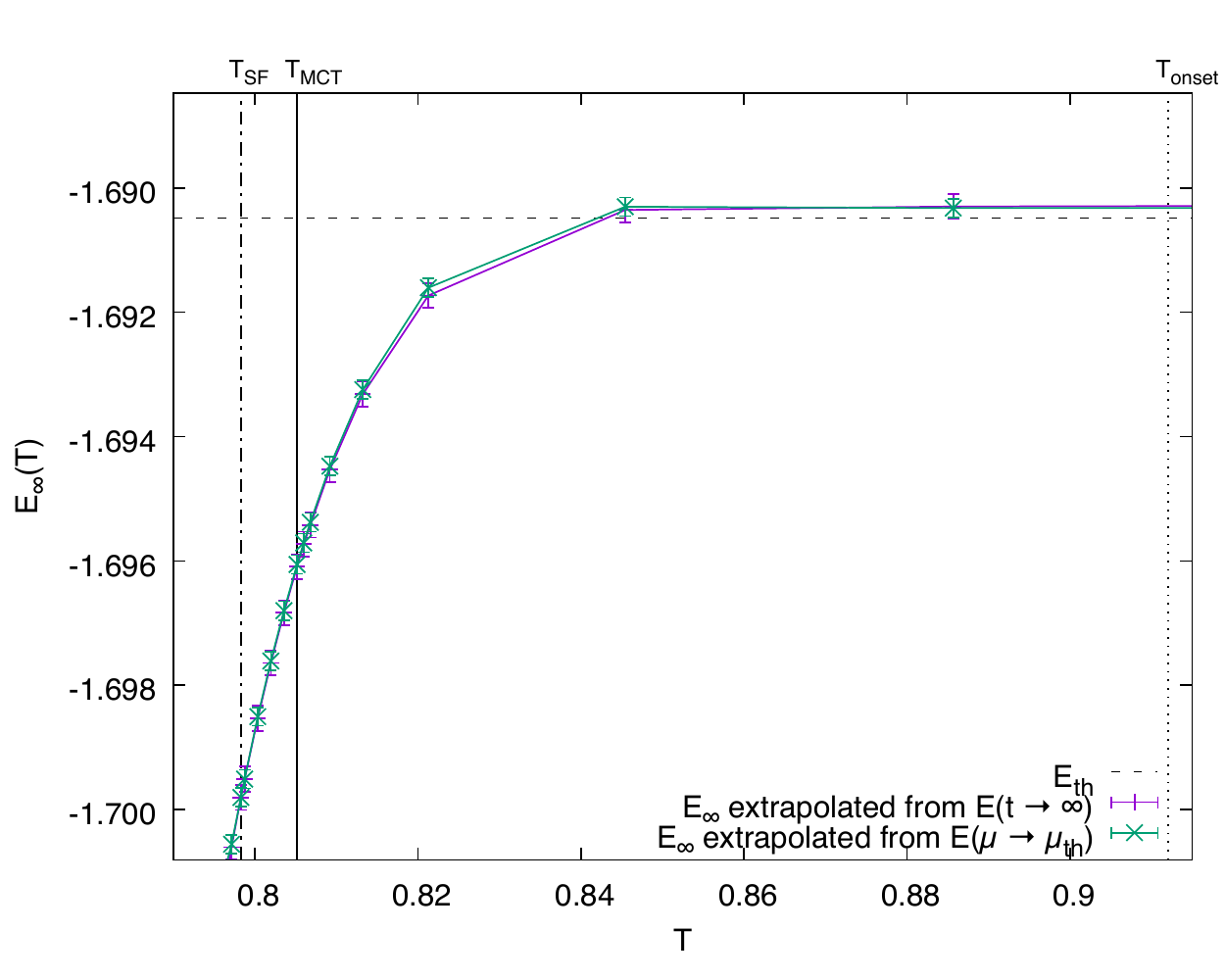}
	\caption{Comparison of two different extrapolations of the asymptotic energies $E_\infty(T)$, obtained fitting versus $t$ and versus $\mu_{mg}-\mu(t)$.}
	\label{fig:2methods} 
\end{figure}

In Fig.~\ref{fig:2methods} we show the two estimates of $E_\infty(T)$ obtained from the procedures just described: the power law fit of energy versus time and the power law fit of energy versus radial reaction $\mu_{mg}-\mu(t)$. We see they are perfectly compatible.

\begin{figure}[t]
	\centering
	\includegraphics[width=0.9\columnwidth]{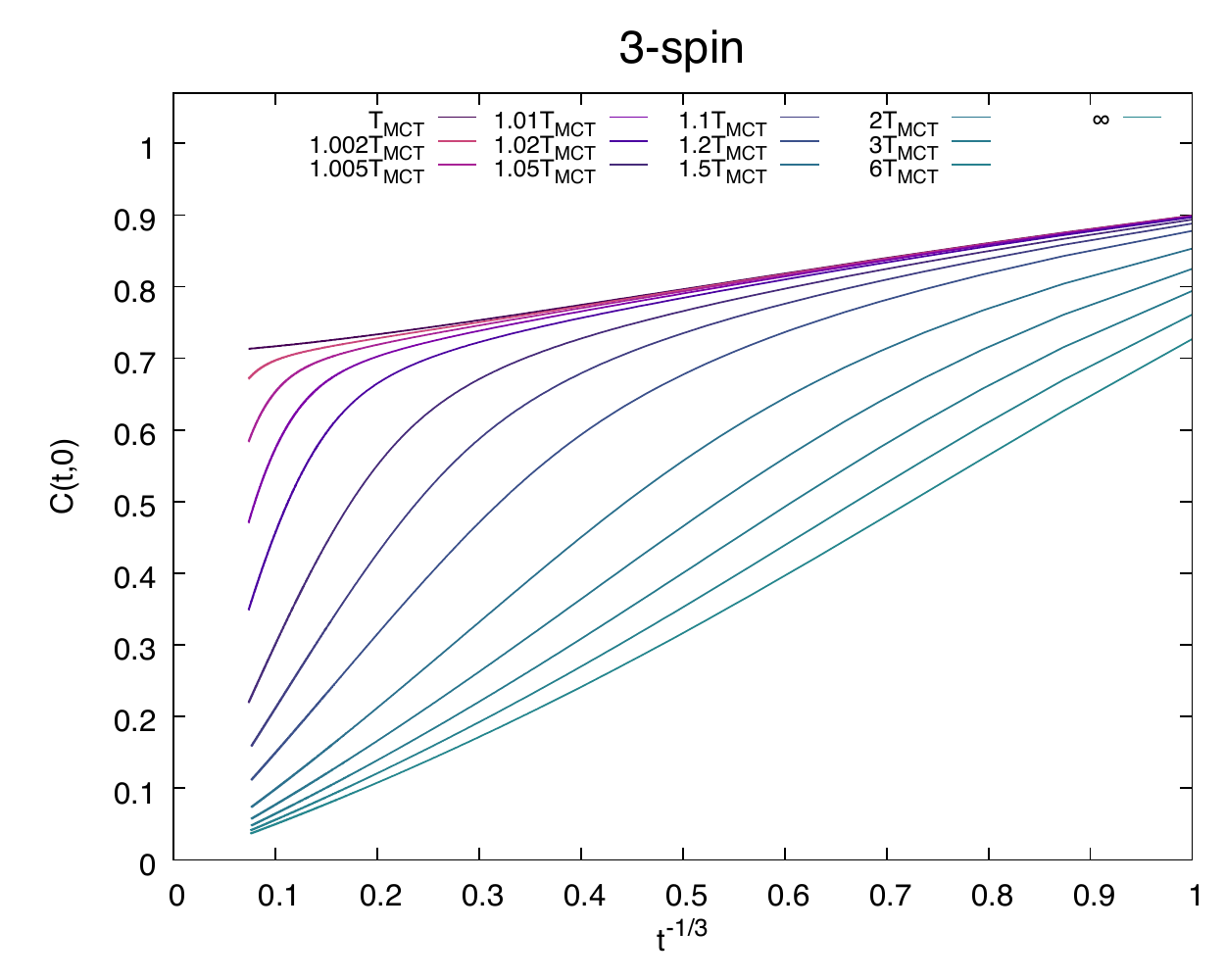}\\
	\includegraphics[width=0.9\columnwidth]{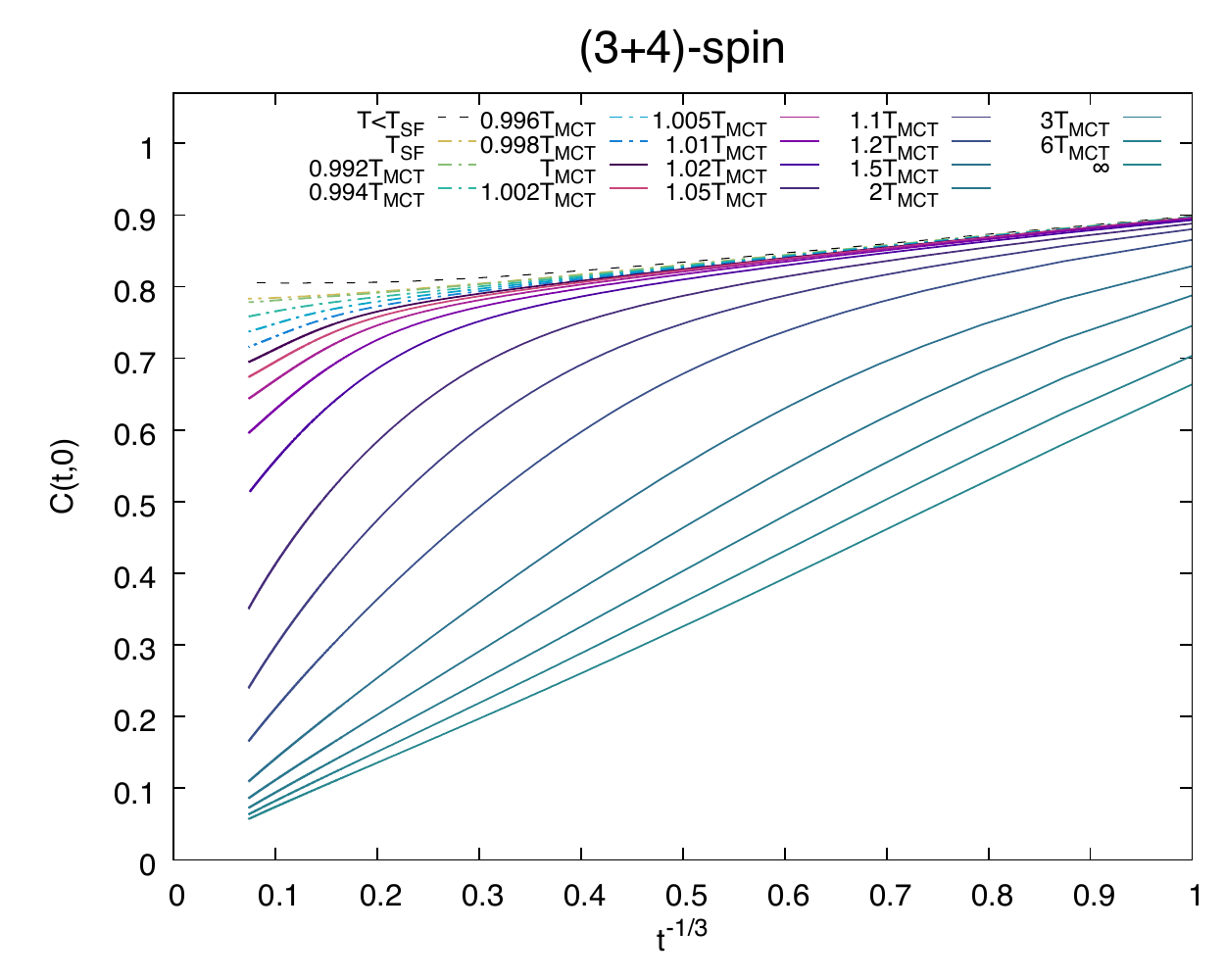}
	\caption{Decay of the correlation with the initial configuration $C(t,0)$ vs $t^{-1/3}$ in the 3-spin model (upper panel) and the (3+4)-spin model (lower panel). Our choice of using the $t^{-1/3}$ variable has been dictated by the request of making the relaxation from $T=\infty$ as linear as possible (notice however that the exact value for the decay exponent of $C(t,0)$ is unknown even in this case). While data for the 3-spin model can be easily extrapolated to zero in the large times limit, the data for the (3+4)-spin model seems much more compatible with a non-zero limit $\lim_{t\to\infty}C(t,0)$ when the temperature gets close to $\TMCT$.}
	\label{fig:correlations} 
\end{figure}

Although we do not want to estimate the large time limit of $C(t,0)$ directly extrapolating in time (in Section~\ref{app:empirical} an estimate will be provided based on a different approach), we believe it is useful to show the $C(t,0)$ curves such that the reader can make her own idea. In Fig.~\ref{fig:correlations} the data for $C(t,0)$ is shown, measured both in the pure 3-spin model and in the mixed (3+4)-spin model. We have decided to show these data as a function of the scaling variable $t^{-1/3}$ that describes pretty well the decay starting from a random configuration ($T=\infty$). En passant, we notice that the exact value for this decay is not known.
We see in Fig.~\ref{fig:correlations} that, while data for the 3-spin model can be easily extrapolated to zero in the large times limit, the data for the (3+4)-spin model seems much more compatible with a non-zero limit $\lim_{t\to\infty}C(t,0)$ when the temperature gets close to $\TMCT$.

The reason why we have not been able to integrate the dynamical equations for a longer time, resides in the difficulties in using integration algorithms with adaptive integration steps. This kind of adaptive algorithms have been used in the past to integrate the dynamical equations in the case of the pure p-spin model starting from random initial conditions \cite{kim_dynamics_2001,berthier_spontaneous_2007}.
Unfortunately, we uncovered that this adaptive algorithm is not robust to changes in the details of the dynamics, and its stability strictly depends on the regime under consideration. At each contraction of the grid, some information on the correlation and response functions computed at the previous steps is lost. This does not seem to be relevant if one starts from a random state and memory of the initial conditions is lost in the asymptotic state. However, in the mixed p-spin model, where memory of the initial configuration crucially affects the large time behavior, the error induced by the time-step adaptation give rise to strong numerical instabilities. Increasing the grid size pushes the instability to later times, but unfortunately, even the largest grid size we could used ($8192 \times 8192$) did not allow substantial improvements with respect to the simple Euler algorithm.

\section{Response at short times}
\label{app:resp_short}

\begin{figure}[t]
	\centering
	\includegraphics[width=\columnwidth]{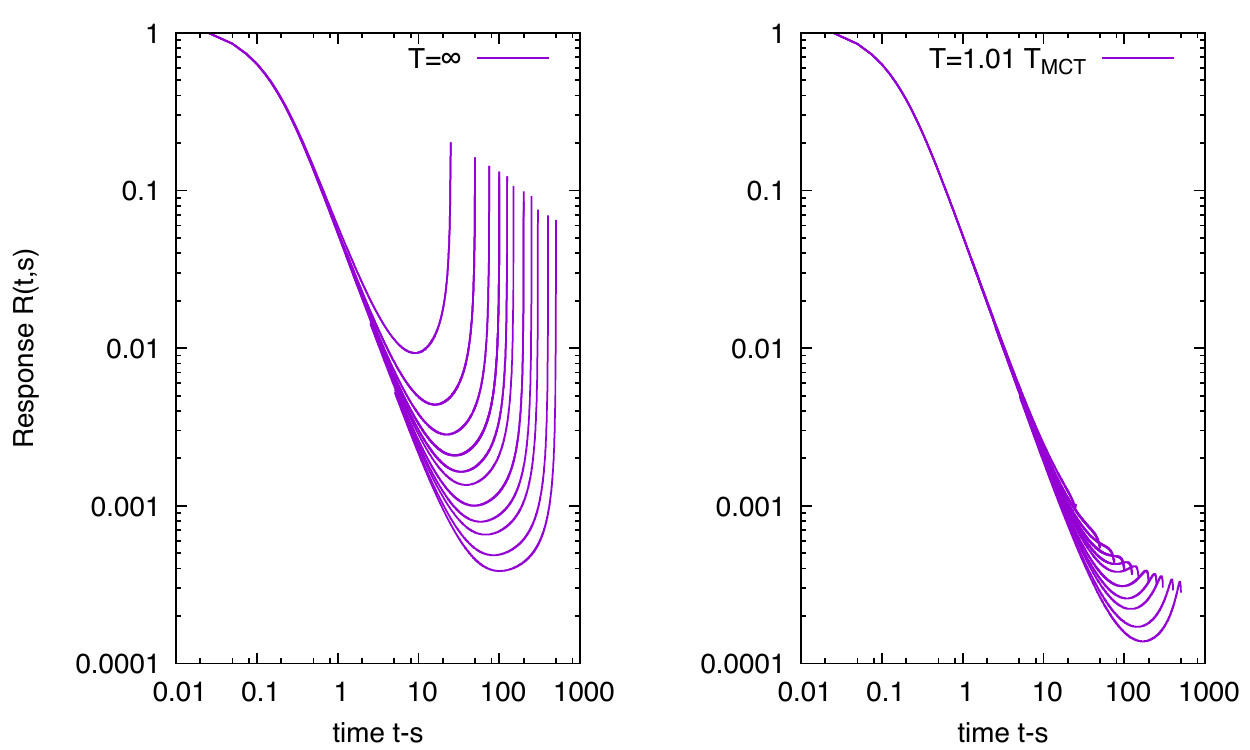}
	\caption{Response function $R(t,s)$ with $T=\infty$ (left) and $T=1.01\,\TMCT$ (right), as a function of the time difference $t-s$ for $t=$ 25, 50, 75, 100, 125, 150, 200, 250, 300, 400 and 500. It is evident that the response is much smaller starting from finite temperature.}
	\label{fig:res}
\end{figure}

\begin{figure}[t]
	\centering
	\includegraphics[width=\columnwidth]{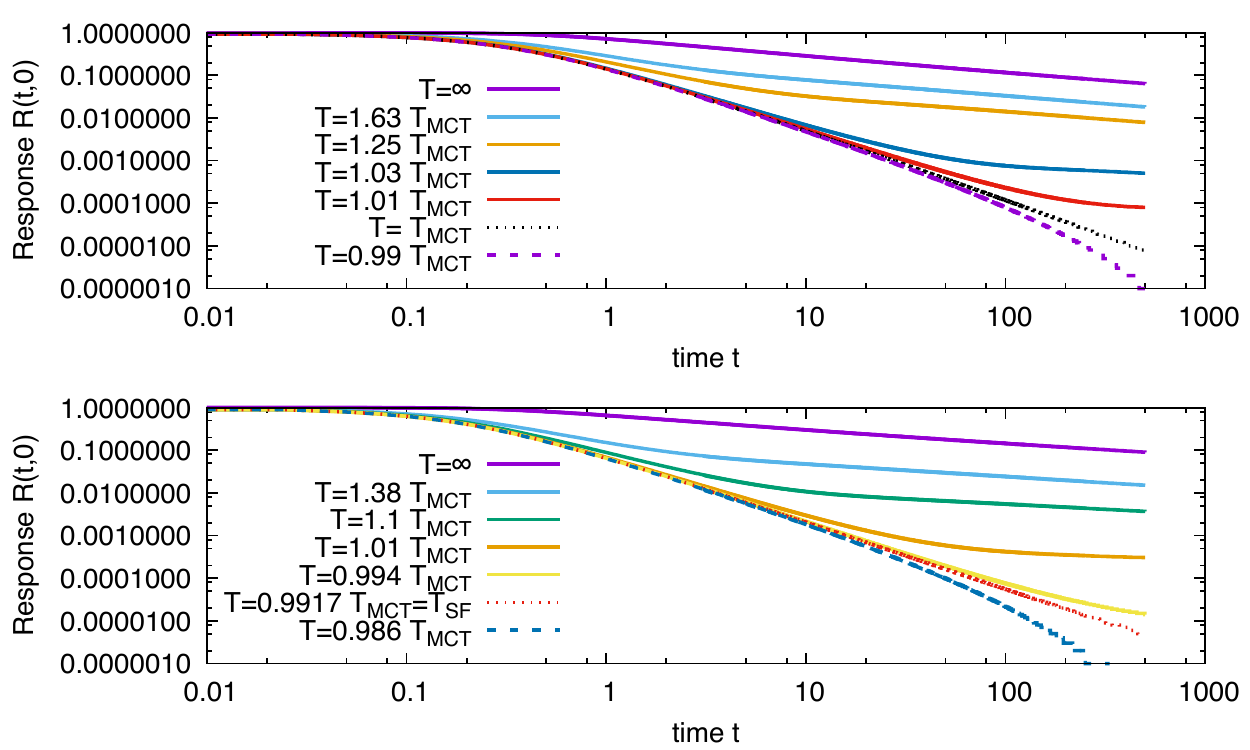}
	\caption{The response to the initial condition $R(t,0)$ for various initial temperatures in the pure 3-spin model (upper panel) and in the mixed 3+4-spin model (lower panel). The qualitative behavior is very similar, the higher is the temperature, the higher are the values of $R(t,0)$. For high temperature we see a slow decay of $R(t,0)$, possibly a power law. At low temperature, below $T_{MCT}$ in the pure model and below $T_{SF}$ in the mixed model, the decay is exponential.
    }
	\label{fig:res2}
\end{figure}

A natural hypothesis about the memory effect in region II is that it could be associated with an increased response $R(t,s)$ to perturbations applied at short times $s$.  Here we show that short time dynamics provides hints against this hypothesis. Let us then compare the response $R(t,s)$ starting from infinite temperature and the one for an initial temperature in the region II. In Fig.~\ref{fig:res} we plot the response as a function of the time difference $t-s$ starting from infinite temperature and from a temperature $T=1.01 \; \TMCT$.  We clearly see that starting from finite temperature the response at short times $s$ is much smaller than the one starting from infinite temperature.
It is also interesting to study the behavior in time of $R(t,0)$, that we plot in Fig.~\ref{fig:res2}. It is well known that starting from a uniformily distributed initial condition one has that the response to the initial condition coincide with the remanent magnetization, namely, 
$R(t,0)=C(t,0)$ \cite{franz_mean_1994}. For finite $T$ this relation does not hold, in fact it is possible to show that $R(t,0)=C(t,0)-\beta \int_0^t ds\; f'(C(0,s))R(t,s)$. We see that a finite limiting value for $C(t,0)$ does not necessarily imply a finite response to the initial condition. Our data show that the two terms largely compensate and the larger $\beta$, the larger the compensation effect.  Both for the mixed and the pure model, the lower the starting temperature, the lower $R(t,0)$. The decay of $R(t,0)$ becomes exponential only for $T=\TSF$. In the pure model this coincide with $\TMCT$, while in the mixed model $\TSF<\TMCT$.

\section{Asymptotic Solutions} 
\label{app:agingRSB}

In Ref.~\cite{barrat_temperature_1997} it was shown that the asymptotic solution to the dynamical equations assuming a simple memorious aging within a state with a unique effective temperature has parameters $\chi$, $q_0$, $q_{12}$ and $y$ that satisfy the same equations that can the obtained extremizing the Franz-Parisi (FP) potential computed in the one step replica symmetry breaking (1RSB) scenario, provided that the parameter $y$, is fixed by a marginality condition.

In our case the 1RSB FP potential has to be computed at zero temperature ($T_f=0$), while the reference configuration is in equilibrium at temperature $T=1/\beta$
\begin{widetext}
\begin{equation}
-2V_\text{\tiny 1RSB}(q_{12},\chi,q_0,y) = \chi f'(1) + y(f(1) - f(q_0)) + \frac1y \log\left(\frac{\chi+y(1-q_0)}{\chi}\right) + \frac{q_0 - q_{12}^2}{\chi+y(1-q_0)} + 2\beta f(q_{12})\;.
\end{equation}
The saddle point equations and the marginality condition thus read
\begin{equation}
\begin{cases}
\hfill \partial_\chi V_\text{\tiny 1RSB} = 0 &\implies \chi(1 - q_{12}^2) + y (1 - q_0)^2 - \chi(\chi + y (1 - q_0))^2 f'(1) = 0
\vspace{2mm}\\
\hfill \partial_{q_0} V_\text{\tiny 1RSB} = 0 &\implies  q_0 - q_{12}^2 - (\chi + y (1 - q_0))^2 f'(q_0) = 0
\vspace{2mm}\\
\partial_{q_{12}} V_\text{\tiny 1RSB} = 0 &\implies q_{12} - \beta (\chi + y (1 - q_0)) f'(q_{12}) = 0
\vspace{2mm}\\
\hfill \text{marginality} &\implies \chi^2 f''(1) = 1
\end{cases}
\label{speRSB}
\end{equation}
\end{widetext}
We notice {\it en passant} that equations in (\ref{speRSB}) do not always select minima of the FP potential, because for that one needs to set to zero the total derivative with respect to $q_{12}$ and not the partial derivative.
The energy and the radial reaction are given respectively by
\begin{eqnarray}
\label{eq:1app}
E &=& - \chi f'(1) - y \big(f(1) - f(q_0)\big) - \beta f(q_{12})\;,\\
\mu &=& \chi f''(1) + (\chi + y) f'(1) - y q_0 f'(q_0) + \beta q_{12} f'(q_{12})\;.\nonumber
\end{eqnarray}

The equations in (\ref{speRSB}) always admit the solution with $q_{12}=q_0=0$, representing the memoryless or `amnesic' aging solution with parameters
\begin{eqnarray*}
\chi&=&\chi_{mg}\equiv\frac{1}{\sqrt{f''(1)}}\;,\\
y&=&y_0\equiv\frac{f''(1)-f'(1)}{f'(1)\sqrt{f''(1)}}=\frac{\sqrt{f''(1)}}{f'(1)}-\chi_{mg}\;,
\end{eqnarray*}
and consequently energy and radial reaction are given by
\begin{equation*}
E=E_{th}\equiv -\chi_{mg} f'(1) - y_0 f(1)\;,
\quad
\mu=\mu_{mg}\equiv 2\sqrt{f''(1)}\;.
\end{equation*}
For the (3+4)-spin model the numerical values for the above parameters are  $\chi_{mg}=1/3$, $y_0=11/21\simeq 0.52381$, $E_{th}=-71/42\simeq-1.69048$ and $\mu_{mg}=6$.

\begin{figure}[t]
	\centering
	\includegraphics[width=\columnwidth]{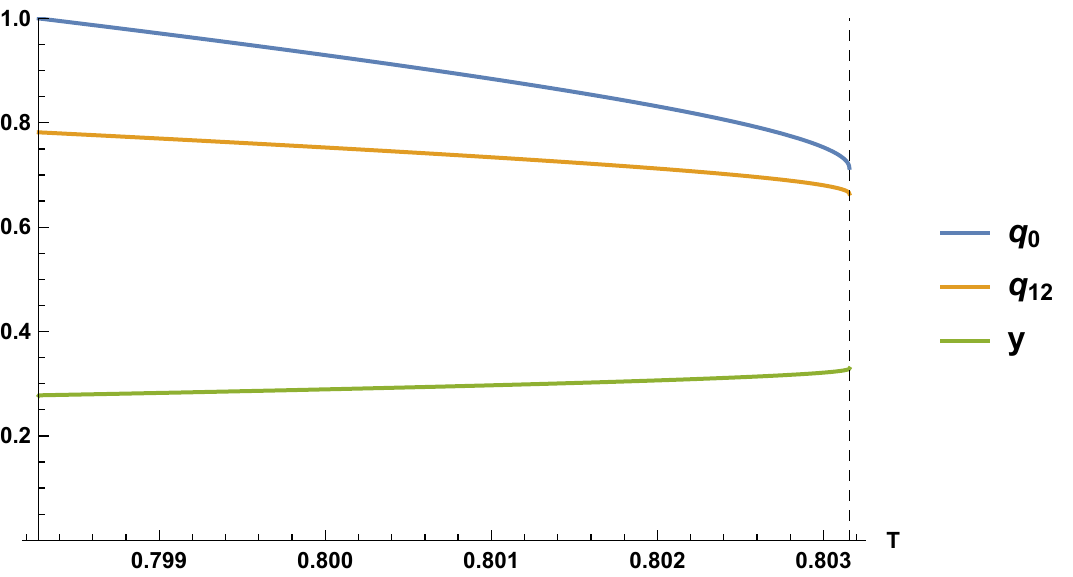}
	\caption{Aging solution with memory of the initial configuration as predicted by the derivative of the FP potential for the (3+4)-spin model. $q_0$ (upper line), $q_{12}$ (central line) and $y$ (lower line) as a function of $T$. The temperature range in the plot is between $\TSF=0.7982754$ and $T_0=0.8031557$ (marked by a dashed vertical line). Notice that in this solution we have $y\approx 0.3$ quite far from the value $y \approx 0.52$ that we observe in the numerical solution to the dynamical equations. Also notice that $q_{12}$ and $q_0$ have finite values when the solution appears at $T_0$.}
	\label{fig:AgSol} 
\end{figure}

A non trivial `memorious' aging solution with strictly positive values for $q_{12}$ and $q_0$, disconnected from the amnesic aging solution, exists only for temperatures below $T_0$. Moreover at an even lower temperature $\TSF$ we have that $q_0\to 1$ and this solution becomes replica symmetric.
In Fig.~\ref{fig:AgSol} we report the values of $q_0$, $q_{12}$ and $y$ in the memorious aging solution.
In the (3+4)-spin model the existence domain for this solution is upper bounded by $T_0=0.8031557$ where the solution disappears by a square root singularity and lower bounded by $\TSF= 0.7982754$, where $q_0\to 1$ and the solution becomes replica symmetric (RS).
At $T=\TSF$ the parameters extremizing the potential can be computed analytically. The value of $q_{12}$ is more easily obtained from the RS solution (see below), while the value of $y$ can be obtained from the third order expansion in $\varepsilon=1-q_0$ of the difference between the two equations in (\ref{speRSB}), thus getting
\begin{equation}
y(\TSF) = \frac{f'''(1)}{2[f''(1)]^{3/2}}
\end{equation}
The numerical integration of the off-equilibrium dynamics does not give any indication in favour of this solution, in particular for the (3+4)-model, this solution has $y\approx 0.3$ in the whole range of validity, which is incompatible with the value $y\approx 0.52$ we obtain from the numerical solution of the dynamics. Even if we were disposed to believe that the dynamics would slowly cross-over to this solution on time scales we cannot observe numerically, still this would not solve the puzzle of the behavior of the asymptotic energy going below threshold for $T>T_0$.  In this solution, the value of $q_0$ tends to 1 at temperature $\TSF$, well above the temperature of the static (Kauzmann) phase transition of the model. The one above is the only exact aging solution that we found; we searched, without success, more complex solutions with more than one effective temperatures.

\begin{figure}[t]
	\centering
	\includegraphics[width=\columnwidth]{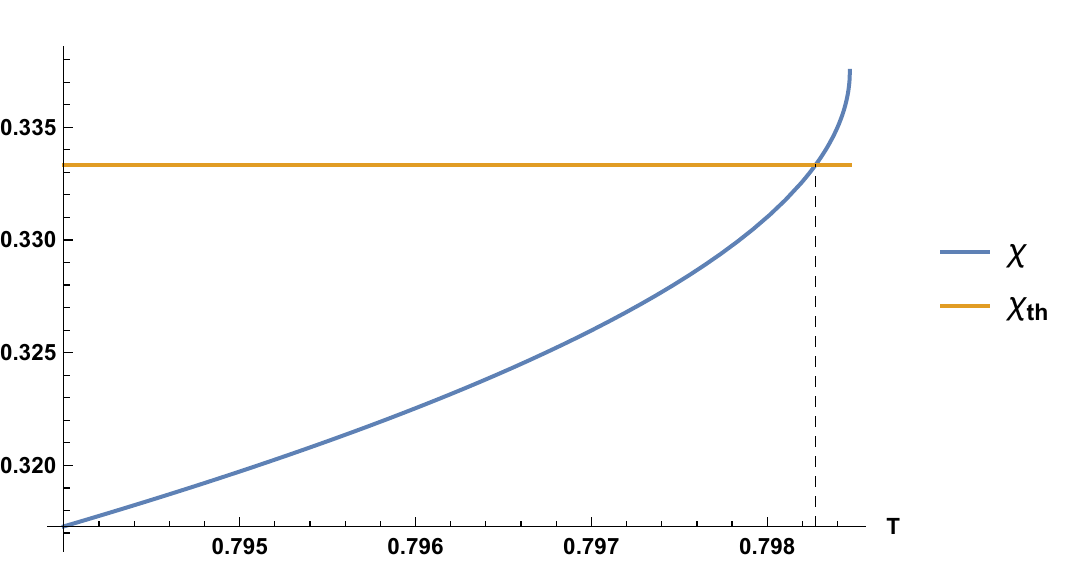}\\
	\vspace{10mm}
	\hfill
	\includegraphics[width=0.9\columnwidth]{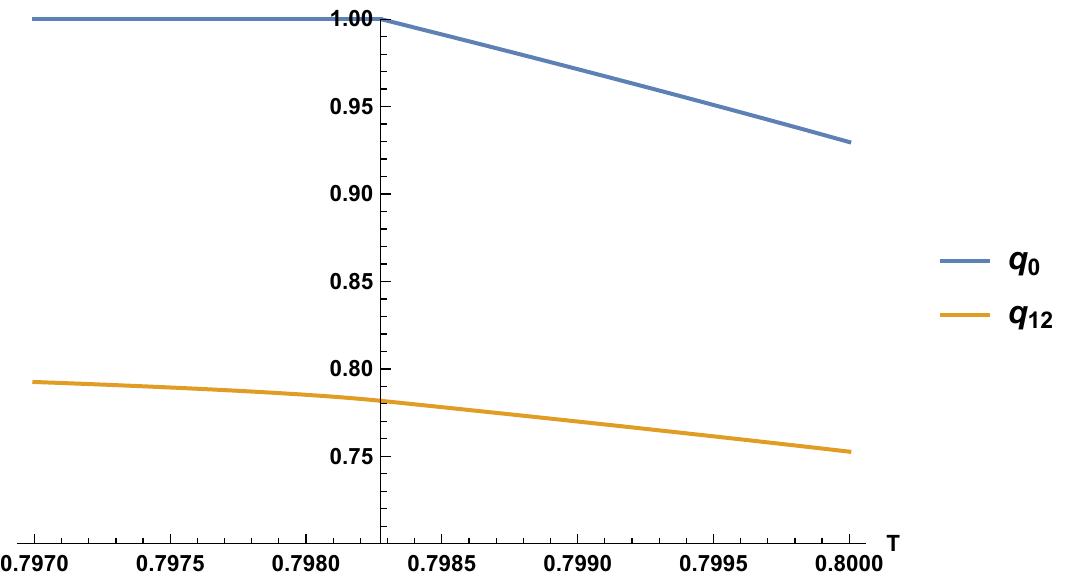}
	\caption{Upper panel: Susceptibility $\chi$ in the RS `state following' solution for the (3+4)-spin model. $\TSF=0.7982756$ is defined from the condition $\chi=\chi_{mg}$, marked by a vertical dashed line.
	Lower panel: Being the vertical axis located exactly at $T=\TSF$, values for $q_0$ and $q_{12}$ on the right are from the RSB aging solution, while those on the left are from the RS `state following' solution.}
	\label{fig:RSsol} 
\end{figure}

\begin{figure*}[t]
	\centering
	\includegraphics[width=\textwidth]{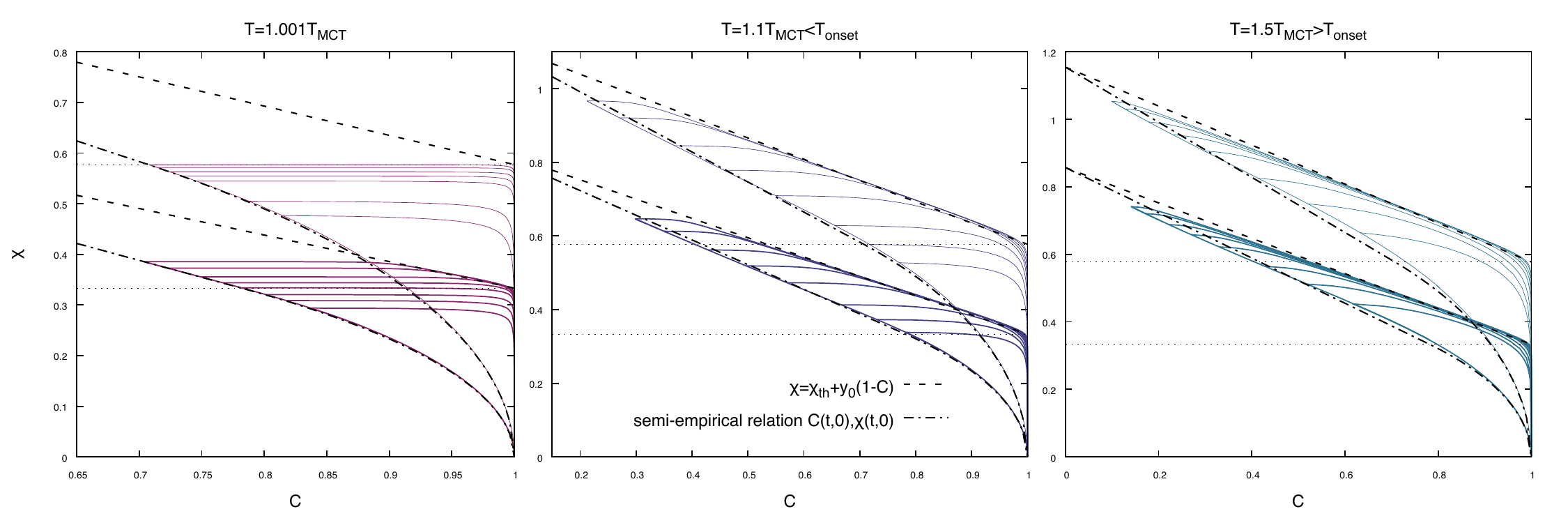}
	\caption{FD curves $\chi(t,s)$ versus $C(t,s)$, each one with a different largest time $t$, plotted parametrically varying the smallest time $s$. In each panel, upper data are for the pure 3-spin model, while lower data are for the mixed (3+4)-spin model. Notice that the 3 panels have different scales, but the dashed and dash-dotted lines are the same in all panels (see text for details).}
	\label{fig:empiricaging}
\end{figure*}

When $q_0\to 1$ the 1RSB potential reduces to the RS potential
\begin{equation}
-2V_\text{\tiny RS}(q_{12},\chi) = \chi f'(1) + \frac{1-q_{12}^2}{\chi} + 2\beta f(q_{12})\;,
\end{equation}
and saddle point equations read
\begin{eqnarray}\label{eq:8}
\begin{cases}
1 - q_{12}^2 - \chi^2 f'(1)=0\\
q_{12} - \chi \beta f'(q_{12})=0\\
\end{cases}
\end{eqnarray}
with the energy taking the value $E=-\chi f'(1)-\beta f(q_{12})$. 
In the 3+4 model, the energy of this solution at $\TSF$ is equal to
$E_\text{\tiny SF}=-1.69981$ and decreases below. 
The susceptibility $\chi$ takes the marginal value $\chi_{mg}$
at $\TSF$ and becomes smaller than this value at lower $T$, as shown in the upper panel of Fig.~\ref{fig:RSsol}.
Our estimate $\TSF=0.7982756$ comes exactly from imposing $\chi=\chi_{mg}$ in this RS solution.
In the lower panel of Fig.~\ref{fig:RSsol} we plot the value of $q_0$ and $q_{12}$ obtained in the RS `state following' solution on the left of the vertical axis, which is located exactly at $T=\TSF$. On the right of the vertical axis the report the corresponding values computed in the aging RSB solution (the same plotted in Fig.~\ref{fig:AgSol}). We observe that $q_{12}$ is a smooth function at $\TSF$. Its value right at $\TSF$ can be computed analytically from the first equation in (\ref{eq:8}) by imposing $\chi=\chi_{mg}$, leading to
\begin{equation}
q_{12,\text{\tiny SF}} = \sqrt{1-\frac{f'(1)}{f''(1)}}\;.
\end{equation}

For $T\le\TSF$ the FP potential has a secondary minimum described by a replica symmetric ansatz. So we are effectively describing the quenching process from temperature $T$ to zero temperature as a `state following' process \cite{sun_following_2012}: the observation that both at the starting and ending temperatures the state we are following is well described by a replica symmetric ansatz is an evidence that static-dynamics equivalence hold in this case. And this is what we indeed observe by comparing the asymptotic dynamics obtained integrating numerically the dynamical equations to the values derived here  from the thermodynamic FP potential. For $T\le\TSF$ aging disappears and one finds a simple agingless relaxation within a state described by the parameters computed in the secondary minimum of the FP potential.

\section{Semi-empirical relations to describe the large times dynamics}
\label{app:empirical}

We have seen that there is no simple aging solution to the asymptotic
equations with $q_{12}>0$ that is consistent with the finite time
numerical integration of the dynamical equations.
Nonetheless the FD plot in the mixed (3+4)-spin model do not differ
in any appreciable way from the usual ones of the pure \pspin model
with a single effective temperature and a value of $y$ equal to the one that holds for the $\beta=0$ initial
condition.  Insisting in representing the best that we can the data
in the {\it hic sunt leones} region II with a single effective temperature
dynamical ansatz, we ask what kind of relation can be derived between the overlaps
$q_0$ and $q_{12}$.
We suppose that $\mu$, $\chi$ take the marginal values $\mu_{mg}\equiv 2\sqrt{f''(1)}$ and $\chi_{mg}\equiv1/\sqrt{f''(1)}$.
Moreover we assume the FD slope in the range $C(t,t')\in[q_0(t),1]$ is independent of the initial temperature, thus it is equal to value obtained with $\beta=0$, i.e.\ $y_0\equiv \sqrt{f''(1)}/f'(1)-1/\sqrt{f''(1)}$.
Finally we assume the response is null in the range $C(t,t')\in[C(t,0),q_0(t)]$, where the correlation decays extremely fast.

Within this 1RSB ansatz the asymptotic value for the radial reaction is given by
\begin{equation*}
\mu  =  \chi f''(1) + (\chi+y) f'(1) - y f'(q_0) q_0 + \beta f'(q_{12}) q_{12}
\end{equation*}
and by imposing $\mu=\mu_{mg}$, $\chi=\chi_{mg}$ and $y=y_0$ we get the relation
\begin{equation}
y_0 f'(q_0) q_0 = \beta f'(q_{12}) q_{12}
\label{eq:main}
\end{equation}
that must hold on the asymptotic solution. We can notice at this point, that using (\ref{eq:1app}) the energy can be written as $E=E_{th}+\Delta E$ with $\Delta E= y f(q_0)- \beta f(q_{12})$.  Only in the pure model Eq.~(\ref{eq:main}) implies $\Delta E=0$, in all the other cases it gives a non vanishing $\Delta E$. 

A second relation between $q_0$ and $q_{12}$ can be derived from the observation of the FD plots in Fig.~\ref{fig:empiricaging}, where each curve is a parametric plot of $\chi(t,t')$ versus $C(t,t')$ at fixed $t$ varying $t'$. We notice that, below the onset temperature, the points $\big(C(t,0),\chi(t,0)\big)$ follows closely the dash-dotted curve that we are going to derive analytically. It is important to remark that the dash-dotted curve is the same in all the panels of Fig.~\ref{fig:empiricaging}, so it is a temperature independent relation. Also the dashed line is temperature independent and corresponds to the FD relation in the aging solution with $\beta=0$, that is a line of slope $-y_0$ connecting the points $(0,\chi_{mg}+y_0)$ and $(1,\chi_{mg})$.

\begin{figure}[t]
	\centering
	\includegraphics[width=\columnwidth]{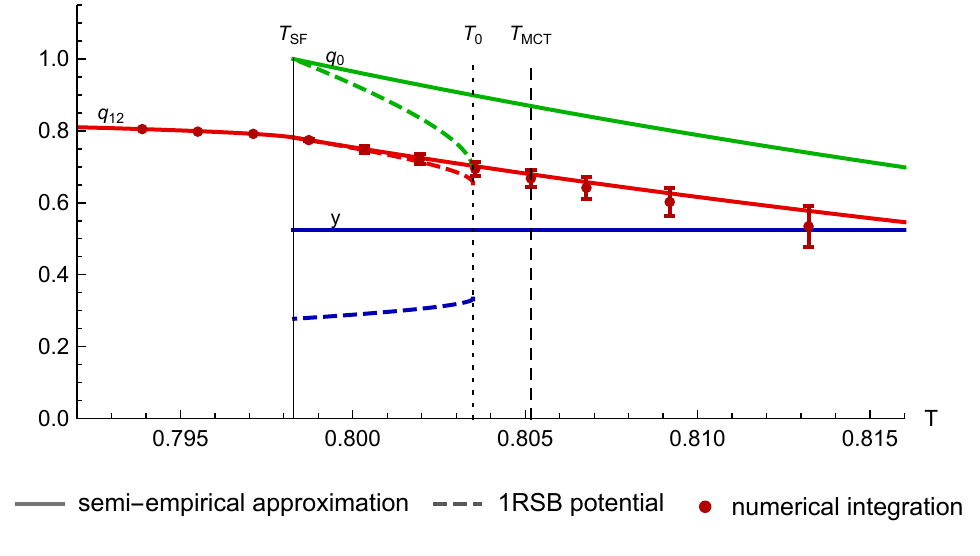}
	\caption{Comparison between asymptotic aging parameters extrapolated from the numerical solution of the dynamical equations (data with errors), the analytical prediction obtained via the semi-empirical approximation derived in App.~\ref{app:empirical} (full curves) and the standard RSB aging solution derived in App.~\ref{app:agingRSB} (dashed curves).}
	\label{fig:qpyplot}
\end{figure}

In order to obtain the dash-dotted line we assume that in the aging regime (i.e.\ for $\chi(t,0)\ge\chi_{mg}$) the relation between $C(t,0)$ and $\chi(t,0)$ is linear, while in the regime $\chi(t,0)<\chi_{mg}$ the dynamics is asymptotically exploring a state and thus we assume the relation that holds within the RS `state following' solution. The latter can be easily derived from the first equation in (\ref{eq:8})
\begin{equation}
\chi = \sqrt{\frac{1-q_{12}^2}{f'(1)}}
\end{equation}
that thus holds for $q_{12}\in[q_{12,\text{\tiny SF}},1]$, while the linear part has slope $-y_0/q_{12,\text{\tiny SF}}$ and passes through the points $(0,\chi_{mg}+y_0)$ and $(q_{12,\text{\tiny SF}},\chi_{mg})$. We notice that the dash-dotted curve has a continuous first derivative at the point $(q_{12,\text{\tiny SF}},\chi_{mg})$, as can be easily checked by taking derivatives.

Our asymptotic ansatz thus implies a very simple relation between the overlaps describing the asymptotic aging regime, namely
\begin{equation}
q_{12} = q_{12,\text{\tiny SF}}\;q_0\;,
\end{equation}
and plugging this relation inside Eq.~(\ref{eq:main}) it is easily to find that a solution with $q_{12}>0$ can exists only if
\begin{equation}
T<\Tonset\equiv\frac{q_{12,\text{\tiny SF}}^k}{y_0}=\frac{f'(1)[f''(1)-f'(1)]^{\frac{k}{2}-1}}{f''(1)^\frac{k-1}{2}}\;,
\end{equation}
where $k$ is defined by $f(q)\propto q^k$ for $q\to 0$  (in the 3+4-model $\Tonset=0.91$). Despite this is not an exact solution of the asymptotic equations, it is a strong indication that there is a phase transition between a memoryless phase where dynamics decorrelates from the initial condition and falls over the `usual' threshold states with $E=E_{th}$ and a phase where aging takes place in a confined space, with an asymptotic energy below threshold and depending on $T$.

In Fig.~\ref{fig:qpyplot} we report an even stronger evidence that this approximate solution provides a very good description of the asymptotic dynamics obtained by numerically integrating the dynamical equations.
We plot numerically extrapolated values as data points with errors, while full lines are prediction from the approximate solution presented in this Appendix and dashed lines correspond to the standard 1RSB aging solution discussed in the previous Appendix. Needless to comment on which one is better in describing the asymptotic dynamics.


\section{Counting the minima}
\label{app:counting}

In this Appendix we try to answer to the question whether the attractors of the dynamics can be well described in terms of typical marginal saddles and minima that lie close to the initial configuration.
Let us consider the stationary points of the Hamiltonian $H[\sigma]$ on the sphere $\sum_i \sigma_i^2=N$:
\begin{eqnarray}
  \label{eq:2app}
 H'_{i} +\mu \sigma_i = 0.
\end{eqnarray}
As in dynamics, the radial reaction $\mu$, takes, in any stationary point,  the value 
\begin{eqnarray}
  \label{eq:3app}
   \mu= -\frac{1}{N}\sum_i \sigma_i H'_{i}.
\end{eqnarray}
We wish to  classify the stationary points according to their energy $E=\frac{1}{N}H[\sigma]$ and the value of the radial reaction $\mu$.

Differently from the pure models where $\mu = p E$, the relation
between $E$ and $\mu$ here is not univocal and stationary points are
found in a whole region of the $(E,\mu)$ plane. 

We have seen that for
$T>\TSF$ dynamics is attracted by some family of marginal minima.
In order to characterize these minima, 
we count the number of stationary points of the Hamiltonian
$H[\vc{\sigma}]$ with fixed energy $E$ and radial reaction $\mu$ that
lie at a fixed overlap $\vc{\sigma}\cdot\vc{\sigma}_0 = N q_{12}$ from a
reference configuration $\vc{\sigma}_0$, weighted with a Gibbs measure
$e^{-\beta H[\vc{\sigma}_0]}$. Since the complexity, i.e.\ the
logarithm of their number, is self-averaging, we write
\begin{widetext}
\begin{eqnarray}
  \label{eq:4}
	\Sigma(E,\mu,q_{12},\beta) =  \int_{S_N} \mathscr{D} \vc{\sigma}_0 \; \frac{e ^ {-\beta H[\vc{\sigma}_0]}}{Z_{\beta}} \; \log\left( \int_S \mathscr{D} \vc{\sigma} \;\delta(Nq_{12}-\vc{\sigma}\cdot\vc{\sigma}_0) \; \delta(NE-{H}) \; \delta(\mu \vc{\sigma}+ {\vc{H}'}) \; |\det (\mu\vc{I}+\vc{H}'')|\right) \qquad
\end{eqnarray}
The computation of $\Sigma$ is standard, and can be performed in
several steps. First of all, since the matrix $\vc{H}''$ is a GOE
random matrix, the distribution of eigenvalues of $\mu\vc{I}+\vc{H}''$
is self-averaging and is a shifted semicircular
$\rho(\lambda) = \frac{2}{\pi 4 f''(1)}\sqrt{4
  f''(1)-(\lambda-\mu)^2}$. The modulus of the determinant
$|\det (\mu\vc{I}+\vc{H}'')|$ is self-averaging and its logarithm
reads
\begin{multline}
D(\mu)=	\frac 1N \log |\det(\mu\vc{I}+\vc{H})| =\\
= \text{\textbf{Re}}\left[ \frac{1}{4 f''(1)} (\mu^2-\mu  \sqrt{\mu ^2-4 f''(1)}-2 f''(1) \left(-2 \log \left(\sqrt{\mu ^2-4 f''(1)}+\mu \right)+1+\log (4)\right)+\mu ^2)\right]
\end{multline}
which only depends on $\mu$. The imaginary part of this expression is the proportion of negative eigenvalues.

To evaluate the remaining terms we use replicas and write $\Sigma \equiv \overline{\log(\mathcal{N})} =  \lim_{n\rightarrow0}\frac{\overline{\mathcal{N}^n}-1}{n}$.
We concentrate to the case of temperatures greater that the static transition temperature ($T>T_K$) of the model 
where the partition function appearing in the denominator of (\ref{eq:4}) is self-averaging and takes its annealed value $\overline{Z_{\beta}} = e^{\frac{N}{2}\beta^2f(1)}$. One can then average over the disorder and the configuration $\sigma_0$ at the same time. 
Opening the delta function in the Fourier basis, 
	\begin{equation}\label{calc}
	\Sigma(E,\mu,p,\beta) =  \lim_{n\to0} \frac{1}{n}\left( e^{-\frac{1}{2}\beta^2f(1)} \int \mathscr{D} s \; e^{\sum^n_a N (i\hat{\beta}_a E-i \hat{\vc{\sigma}}_a\cdot \vc{\sigma}_a\mu)} \;\delta(Nq_{12}-\vc{\sigma}_a\cdot\vc{\sigma}_0) \; \overline{e ^ {-\beta {H}_0} \; e^{\sum^n_a (i\hat{\beta}_a +i \hat{\vc{\sigma}}_a\cdot \vc{\nabla}){H}_a}}\right)+N D(\mu)
	\end{equation}
	where $\int \mathscr{D} s = \int_S \mathscr{D} \vc{\sigma}_0  \prod_a \left( \int_S \mathscr{D} \vc{\sigma}_a \int \mathscr{D} \hat{\vc{\sigma}}_a\int \hat{\beta}_{a}\right)$. 
	
	And since the disorder is Gaussian :
		\begin{equation*}
	 \overline{e ^ {-\beta {H}_0} \quad e^{\sum^n_a (i\hat{\beta}_a +i \hat{\vc{\sigma}}_a\cdot \vc{\nabla}){H}_a}} = e^{\frac{1}{2}\left(\beta^2f\big ( \tfrac{\vc{\sigma}_{0}\cdot\vc{\sigma}_{0}}{N}\big ) + 2\beta\sum_a (i\hat{\beta}_{a} +i \hat{\vc{\sigma}}_{a}\cdot \vc{\nabla}^{a}) f\big ( \tfrac{\vc{\sigma}_{a}\cdot \vc{\sigma}_{0}}{N}\big ) +\sum_{ab} (i\hat{\beta}_{a} +i \hat{\vc{\sigma}}_{a}\cdot \vc{\nabla}^{a})(i\hat{\beta}_{b} +i \hat{\vc{\sigma}}_{b}\cdot \tilde{\vc{\nabla}}^{b})f\big ( \tfrac{\vc{\sigma}_{a}\cdot\tilde{\vc{\sigma}}_{b}}{N}\big )\big|_{\tilde{\vc{\sigma}}\rightarrow\vc{\sigma}} \right)}
	\end{equation*}
	Now we define overlap variables $N Q_{ab} = \vc{\sigma}_{a}\cdot\vc{\sigma}_{b}$, $N \chi_{ab} = i\vc{\sigma}_{a}\cdot\hat{\vc{\sigma}}_{b}$ and $N R_{ab} = -\hat{\vc{\sigma}}_a\cdot\hat{\vc{\sigma}}_b$, and the overlaps with the reference configuration $N q_{12} =  \vc{\sigma}_{a}\cdot\vc{\sigma}_{0}$,  $N\chi_p = i\hat{\vc{\sigma}}_a\cdot\vc{\sigma}_{0}$. This change of variables defines a matrix:
	\[
	\mathscr{Q} \equiv \begin{pmatrix}  
	1&q_{12}&-i\chi_p\\
	q_{12}&Q_{ab} &-i\chi_{ab}\\
	-i\chi_p &-i\chi_{ab} &-R_{ab}\\
	\end{pmatrix}
	\]
	where $Q_{aa} = 1$ due to the spherical constraint. And, from the equivalence between replicas, we fix $i\hat{\beta}_a = y$ and $\chi_{aa} = \chi$ $\forall a$.
With this change of variables Eq.~(\ref{calc}) becomes
	\begin{align}
		\Sigma=&\left( y E - \mu \chi\right)
			+\beta\left(  y  f(q_{12}) +\chi_p f'(q_{12})\right)
			+\lim_{n\to0} \frac{1}{n}\left(\frac{1}{2}\log (\det \mathscr{Q})\right)+D(\mu)\nonumber\\
			&+\lim_{n\to0} \frac{1}{n}\left( \frac{1}{2}\sum_{ab}[y^2 f(Q_{ab})+2 yf'(Q_{a b})\chi_{a b}+ f'(Q_{ab})R_{ab}+f''(Q_{a b})(\chi_{ab})^2]\right)
			\label{general1}
	\end{align}
	where $\frac{1}{2}\log (\det \mathscr{Q})$ is the volume factor that comes from the change of variables from spins to overlaps.
	To get the leading $N$ contribution we must extremize with respect to all the overlap parameters $\mathscr{Q}$ and $y$. We notice that a further simplification to the expression ($\ref{general1}$) can be obtained by firstly extremizing with respect to $R_{ab}$. 
Assuming a replica symmetric ansatz for the overlap matrices $Q$ and $\chi$, i.e. $Q = \delta_{ab} + (1-\delta_{ab})q_0$ and $\chi_{ab} = \delta_{ab}\chi + (1-\delta_{ab})\chi_1$ we get, in the limit $n\rightarrow0$:
	\begin{equation}\label{general2}
		\begin{aligned}
			\Sigma(E,\mu,q_{12},\beta;y,\chi,\chi_1,\chi_p,q_0) = 
			&+y E - \mu \chi
			+\beta [y  f(q_{12}) + \chi_p f'(q_{12})]\\
			&+\frac{1}{2}[y^2 (f(1)-f(q_0))+2y(f'(1)\chi+f'(q_0)\chi_1)+\mathscr{R}+(\chi^2f''(1)-\chi_1^2f''(q_0))]\\
			&+\frac{1}{2}(\log(1-q_0)+\frac{q_0-q_{12}^2}{1-q_0}-\log(f'(1)-f'(q_0))-\frac{f'(q_0)}{f'(1)-f'(q_0)}
			+ D(\mu)
		\end{aligned}
	\end{equation}
	where
	\begin{equation}
	\mathscr{R} \equiv 1+f'(1)(\tfrac{\chi-\chi_1}{1-q_0}((\chi-\chi_1)-\tfrac{q_0-q_{12}^2}{1-q_0}+2(\chi_1-p\chi_p))+\chi_p^2)-f'(q_0)(\tfrac{\chi-\chi_1}{1-q_0}(-\tfrac{q_0-q_{12}^2}{1-q_0}+2(\chi_1-p\chi_p))+\chi_p^2)
	\end{equation}
	This can be extremized explicitly with respect to $y,\chi,\chi_1,\chi_p$, while $q_0$-extremization has to be done numerically. 
For $q_{12}=0$ the solution is $q_0=0$ and we recover the expression found in \cite{arous_geometry_2018}
        \begin{eqnarray}
          \label{eq:9app}
\Sigma(E,\mu)&=&\max\left\{0,   {\rm Re}\;  \left(-\frac{ \left(E^2
   \left(f''(1)+f'(1)\right)+2 E\mu 
   f'(1)+f(1) \mu ^2\right)}{2 f(1)
   \left(f''(1)+f'(1)\right)-f'(1)^2}
\right. 
\right.
\nonumber
\\
&&
\left. 
\left. 
+\frac{ \mu
   }{\sqrt{\mu ^2- 4f''(1)}+\mu }+ \log
   \left(\sqrt{\mu ^2-4 f''(1)}+\mu \right)+\frac 1 2 \log
   \left(\frac{1}{f'(1)}\right)- \log
   (2)\right)\right\}
        \end{eqnarray}
\end{widetext}
This unconstrained complexity is plotted in  Fig.~\ref{fig:Complexity}. For any value of $\mu$, $\Sigma(E,\mu)$ has a concave parabolic shape as a function of the energy and many values of the energy are possible for the same $\mu$ value. In particular this is true for $\mu=\mu_{mg}$: there is not a unique threshold energy (at variance to pure \pspin models), but a whole interval for which $\Sigma(E,\mu_{mg})>0$. We define the dominating stationary points at a given value of the energy as the ones that maximize $\Sigma$ as a function of $\mu$.  The threshold energy as defined from dynamics, corresponds to the values of the energy that separates minima from saddles on the dominating line. Notice that this value does not correspond to the most numerous marginal minima, which occurs for $\mu=\mu_{mg}$ and $E>E_{th}$, but to the point where the most numerous stationary saddles become minima. This observation sheds some light on the dynamics from a random initial condition; minima are not accessible at levels of the energy $E>E_{th}$ where saddles dominate the landscape, even if many minima are present.

\begin{figure}[t]
	\centering
	\includegraphics[width=\columnwidth]{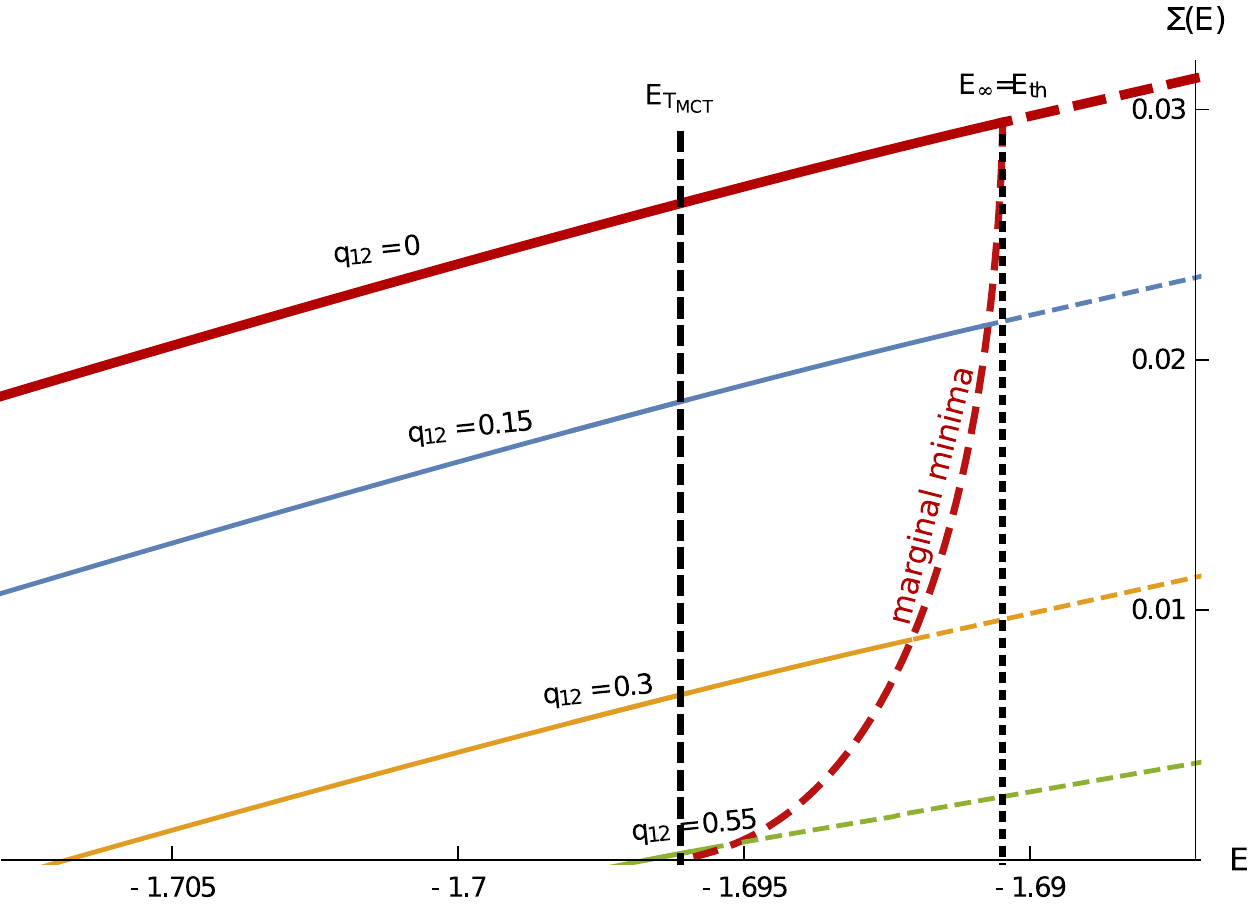}
	\caption{
	Constrained complexity at an overlap $q_{12}$ from a reference configuration sampled at temperature $T=\TMCT$. Vertical lines mark energy values $E_{\infty}$ (dotted line on the right) and $E_{\TMCT}$ (dashed line on the left), corresponding to extrapolated asymptotic energies reached by the dynamics starting respectively from $T=\infty$ and $T=\TMCT$. From a random configuration with $T=\infty$ the dynamics goes to the energy level $E_{th}$, where the dominant stationary points are marginal minima, while starting near $\TMCT$ the dynamics goes below $E_{th}$. If $q_{12}>0$ the energy where marginal minima dominate decreases and we represent it by a bold dashed red curve.}
	\label{fig:Complexityapp}
\end{figure} 

Extending the concept of threshold energy to the case $q_{12}>0$ is straightforward. As shown in  Fig.~\ref{fig:Complexityapp} for any fixed value of $q_{12}$ we can plot the curve corresponding to dominating minima. This curve ends at an energy value that we call $E_{th}(q_{12})$ and this is the best candidate for the asymptotic energy in the relaxation dynamics. Indeed for $E>E_{th}(q_{12})$ dominating stationary points are saddle and with high probability the dynamics will be able to relax further until the energy level $E_{th}(q_{12})$ is reached.
However, even assuming the dynamics converges to $E_{th}(q_{12})$, we are still left with an unknown: the value of $q_{12}$, i.e. the asymptotic value for $C(t,0)$.
This is the same problem the authors of Ref.~\cite{capone_off-equilibrium_2006} faced after computing the constrained complexity for the free-energy at a non-zero temperature.

\begin{figure*}[t]
	\centering
	\includegraphics[width=\columnwidth]{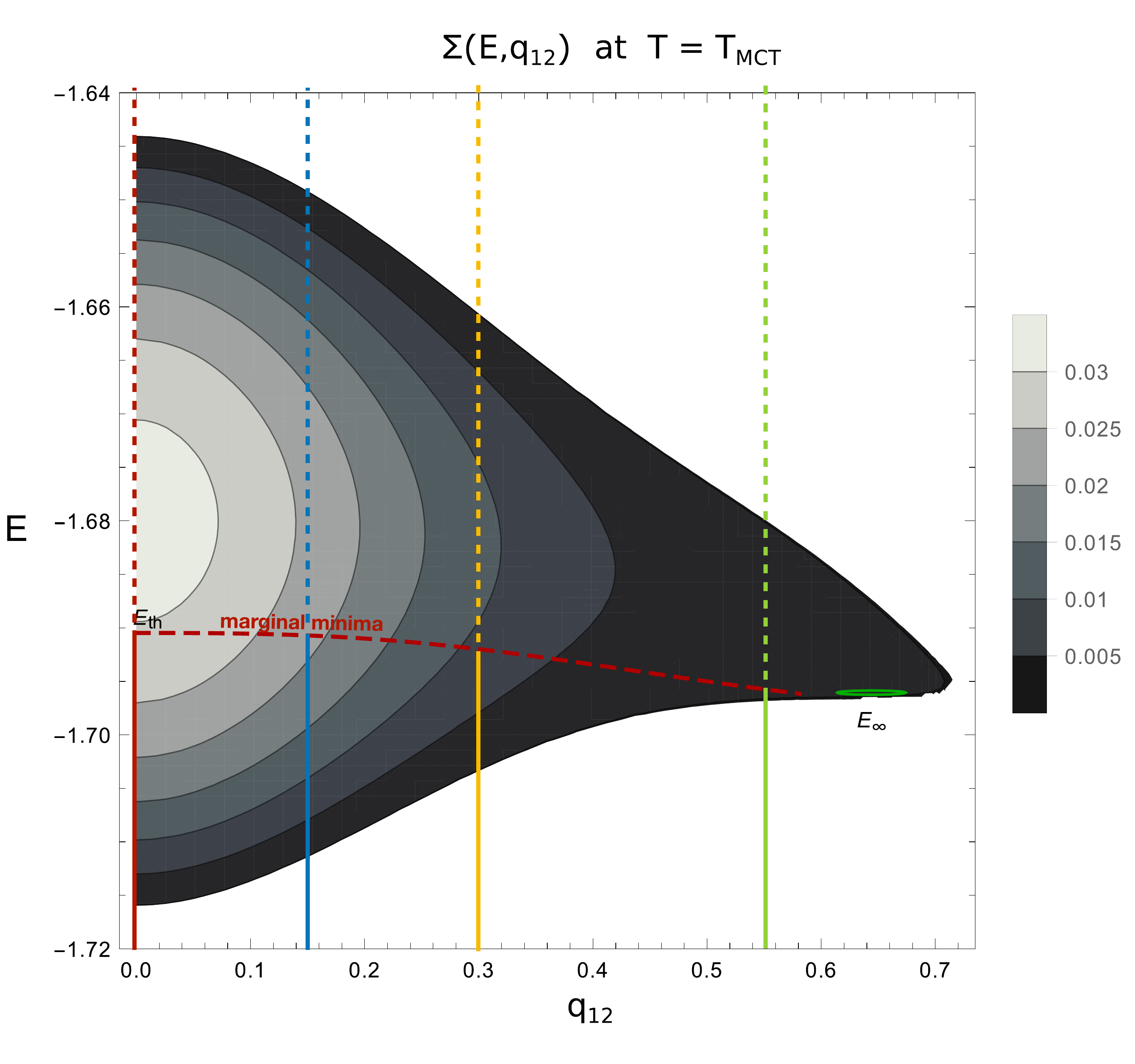}
	\hfill
	\includegraphics[width=\columnwidth]{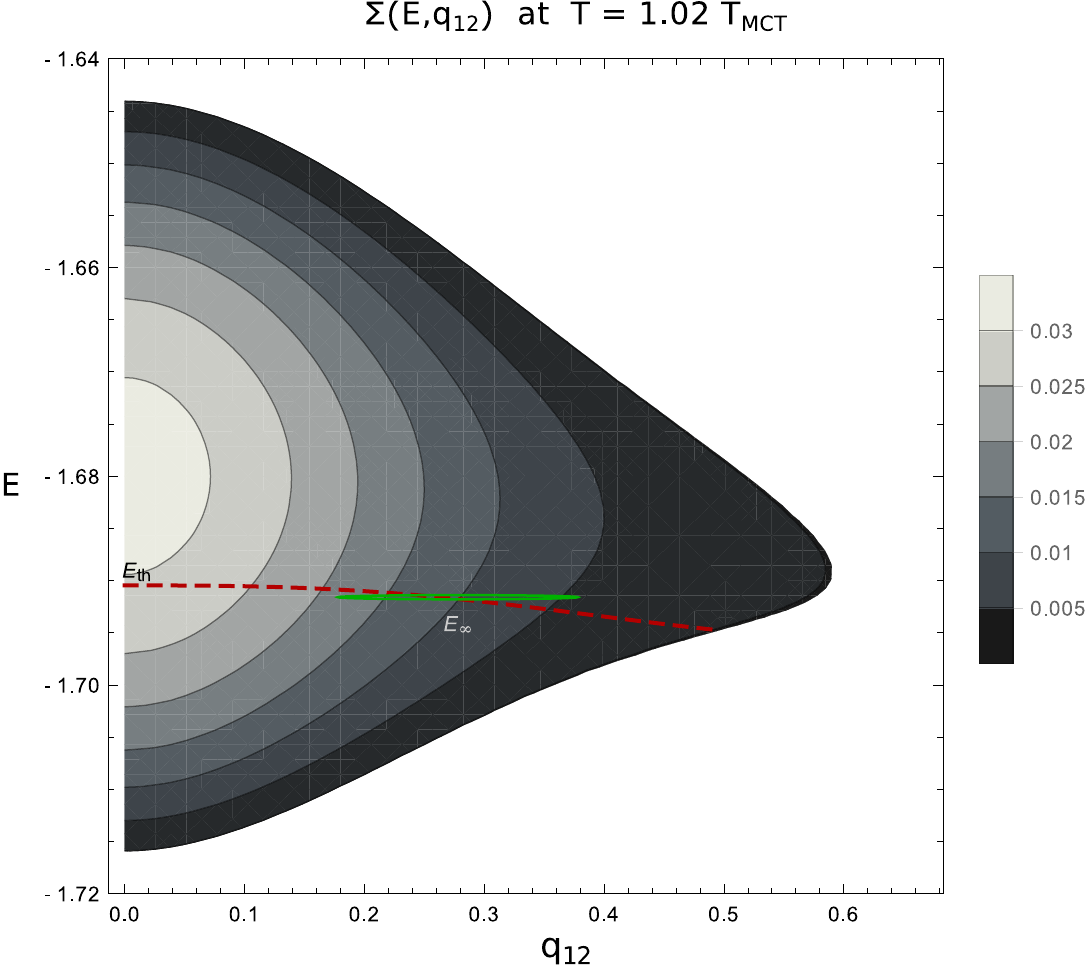}
	\caption{The quenched complexity of dominant states with energy $E$ and overlap $q_{12}$ with respect to an equilibrium configuration at temperature $T=\TMCT$ (left panel) and $T=1.02\,\TMCT$ (right panel). In the left panel the four coloured vertical lines highlight the complexities shown in Fig.~\ref{fig:Complexityapp}. The choice of the temperature is not crucial as long as $T\in(\TSF,\Tonset)$. The almost horizontal red curves mark the threshold energy $E_{th}(q_{12})$ and the green ellipses is our best estimate for the large time limit of the relaxation dynamics obtained from the numerical integration.}
	\label{fig:MargComplexity}
\end{figure*}

In Fig.~\ref{fig:MargComplexity} we plot the quenched constrained complexity of dominant states in the (3+4)-spin model, computed with $T=\TMCT$ (left panel) and $T=1.02\,\TMCT$ (right panel). Please notice that the choice of the temperature is not crucial: as long as $T\in(\TSF,\Tonset)$ the plot would be very similar.
The almost-horizontal red curve in the plot represents the threshold energy $E_{th}(q_{12})$ defined above.
We notice that the range of energies with a positive complexity is very large compared to variations in $E_{th}$.
The green ellipses represents our best estimate for the large time limit of the actual dynamics solved numerically: while the energy can be very well estimated, the limit of $C(t,0)$ is plagued by a large uncertainty due to its slow convergence (until now we have not attempted such an extrapolation, but here we want to be more speculative and we take the risk).
From the plot one is tempted to conjecture that the dynamics always converges to marginal states with a threshold energy $E_{th}(q_{12})$.
However, we have not found any principle to fix the value of $q_{12}$ solely from the complexity curve and further studies are needed to better match the large time limit of the dynamics to the energy landscape.

\begin{figure}[t]
	\centering
	\includegraphics[width=\columnwidth]{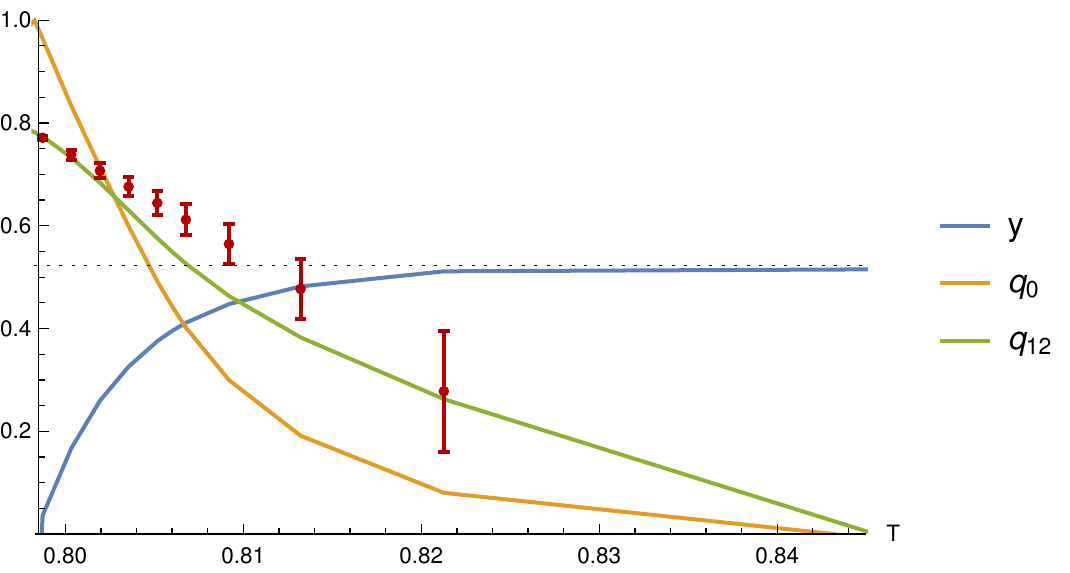}
	\caption{Assuming the dynamics relaxes on the marginal manifold with energy $E_{th}(q_{12})$ and fixing the energy from the large time extrapolation of the numerical data, which in turn fixes the value for $q_{12}$, we can compute analytical values for the remaining aging parameters, namely $q_0$ and $y=\partial_E \Sigma$. While the estimate of $q_{12}$ is compatible with the large time extrapolation of $C(t,0)$ (red points with error in the figure), the other aging parameters are far from those measured in the actual dynamics.}
	\label{fig:yqp}
\end{figure}

Although Fig.~\ref{fig:MargComplexity} may suggest a relation between dynamics and the generalized threshold energies, a more careful analysis reveals its limitations.
Assuming that at large times the relaxation dynamics converges to the manifold of marginal states belonging to the curve $E_{th}(q_{12})$ one could estimate the point reached by the dynamics by extrapolating the asymptotic energy $E_\infty(T)$ and estimating $q_{12}$ from the equality $E_{th}(q_{12})=E_\infty$. Having thus fixed the values of $E$ and $q_{12}$ one can proceed estimating the remaining parameters of the asymptotic aging dynamics, $q_0$ and $y=\partial_E \Sigma$, from the saddle point equations used to compute the quenched complexity. The result of this computation is shown in Fig.~\ref{fig:yqp} with full lines, and compared to the (very uncertain) extrapolation of $C(\infty,0)$, shown by red points with errors. We clearly see that, while the estimate of $q_{12}$ is compatible with the actual dynamics, the other two parameters are far from the values measured in the numerical solution to the dynamics. Indeed $q_0$ becomes smaller than $q_{12}$ while in the actual dynamics the inequality $q_{12}<q_0$ is always satisfied, and $y$ becomes much smaller than $y_0$ (marked by a dotted horizontal line in Fig.~\ref{fig:yqp}) which is good descriptor of the actual aging in the whole temperature range studied. So, the present computation of the quenched complexity of marginal minima of the energy function in mixed \pspin models does not allow to identify the attractors of the dynamics. More work will be required to connected the large time aging dynamics to the properties of the energy landscape.

\bibliography{biblio}

\end{document}